\documentclass[sn-mathphys]{sn-jnl}
\usepackage{graphicx}
\usepackage[separate-uncertainty = true]{siunitx}
\graphicspath{{img/}}

\usepackage{physics}
\usepackage{amsmath}
\usepackage{color}
\usepackage{times}
\usepackage{hyperref}
\usepackage{comment}
\usepackage[left]
{lineno}

\begin{document} 

\baselineskip18pt

\title[Far-field Petahertz Sampling of Plasmonic Fields]{Far-field Petahertz Sampling of Plasmonic Fields} 
\author[1,2]{\fnm{Kai-Fu} \sur{Wong}}\email{kai-fu.wong@cfel.de}
\equalcont{These authors contributed equally to this work.}
\author[3,4]{\fnm{Weiwei} \sur{Li}}\email{weiwei.li@physik.uni-muenchen.de}
\equalcont{These authors contributed equally to this work.}
\author[3,4]{\fnm{Zilong} \sur{Wang}}\email{zilong.wang@physik.uni-muenchen.de}
\author[2]{\fnm{Vincent} \sur{Wanie}}\email{vincent.wanie@desy.de}
\author[2]{\fnm{Erik} \sur{M\r{a}nsson}}\email{erik.maansson@desy.de}
\author[1]{\fnm{Dominik} \sur{Hoeing}}\email{dominik.hoeing@uni-hamburg.de}
\author[3,4]{\fnm{Johannes} \sur{Blöchl}}\email{johannes.bloechl@physik.uni-muenchen.de}
\author[3,4]{\fnm{Thomas} \sur{Nubbemeyer}}\email{thomas.nubbemeyer@physik.uni-muenchen.de}
\author[5]{\fnm{Abdallah M.} \sur{Azzeer}}\email{azzeer@ksu.edu.sa}
\author[2,6]{\fnm{Andrea} \sur{Trabattoni}}\email{andrea.trabattoni@desy.de}
\author*[1,7]{\fnm{Holger} \sur{Lange}}\email{holger.lange@uni-hamburg.de}
\author*[1,2]{\fnm{Francesca} \sur{Calegari}}\email{francesca.calegari@desy.de}
\author*[3,4,8,9]{\fnm{Matthias F.} \sur{Kling}}\email{kling@stanford.edu}
\affil[1]{\orgname{The Hamburg Centre for Ultrafast Imaging}, \orgaddress{\street{Luruper Chaussee 149}, \postcode{22761} \city{Hamburg}, \country{Germany}}}
\affil[2]{\orgdiv{Center for Free-Electron Laser Science CFEL}, \orgname{Deutsches Elektronen-Synchrotron DESY}, \orgaddress{\street{Notkestr. 85}, \postcode{22607} \city{Hamburg}, \country{Germany}}}
\affil[3]{\orgname{Max Planck Institute of Quantum Optics}, \orgaddress{\street{Hans-Kopfermann-Str. 1}, \postcode{85478} \city{Garching}, \country{Germany}}}
\affil[4]{\orgname{Physics Department, Ludwig-Maximilians-Universität Munich}, \orgaddress{\street{Am Coulombwall 1}, \postcode{85748} \city{Garching}, \country{Germany}}}
\affil[5]{\orgname{Attosecond Science Laboratory, Physics and Astronomy Department, King-Saud University}, \orgaddress{\street{P.O. Box 2455}, \city{Riyadh} \postcode{11451}, \country{Saudi Arabia}}}
\affil[6]{\orgname{Institute of Quantum Optics, Leibniz Universität Hannover}, \orgaddress{\street{Welfengarten 1}, \postcode{30167} \city{Hannover}, \country{Germany}}}
\affil[7]{\orgname{Institute of Physics and Astronomy, Universit\"at Potsdam}, \orgaddress{\street{Karl-Liebknecht-Str. 24}, \postcode{14476} \city{Potsdam}, \country{Germany}}}
\affil[8]{\orgdiv{Stanford PULSE Institute}, \orgname{SLAC National Accelerator Laboratory}, \orgaddress{\street{2575 Sand Hill Rd}, \city{Menlo Park, CA} \postcode{94025}, \country{USA}}}
\affil[9]{\orgdiv{Applied Physics Department}, \orgname{Stanford University}, \orgaddress{\street{348 Via Pueblo}, \city{Stanford, CA} \postcode{94305}, \country{USA}}}


\abstract{
The collective response of metal nanostructures to optical excitation leads to localized plasmon generation with nanoscale field confinement driving applications in e.g. quantum optics, optoelectronics, and nanophotonics, where a bottleneck is the ultrafast loss of coherence by different damping channels. The present understanding is built-up on indirect measurements dictated by the extreme timescales involved. Here, we introduce a straightforward field sampling method that allows to measure the plasmonic field of arbitrary nanostructures in the most relevant petahertz regime. We compare experimental data for colloidal nanoparticles to finite-difference-time-domain calculations, which show that the dephasing of the plasmonic excitation can be resolved with sub-cycle resolution. Furthermore, we observe a substantial reshaping of the spectral phase of the few-cycle pulse induced by this collective excitation and we demonstrate ad-hoc pulse shaping by tailoring the plasmonic sample. The results pave the way towards both a fundamental understanding of ultrafast energy transformation in nanosystems and practical applications of nanostructures in extreme scale spatio-temporal control of light.}


\maketitle

\section*{Introduction}
In metallic nanoparticles (NPs) the light electric field can drive the conduction band electrons into a collective oscillation on the nanoscale, referred to as localized surface plasmon (LSP) \cite{maier_plasmonics_2007}. Coupling of light to a LSP resonance leads to the local enhancement of the electromagnetic field and to the confinement of the light-matter interaction on the nanoscale, therefore enabling a manifold of applications including surface enhanced spectroscopy \cite{schlucker_surface-enhanced_2014}, enhanced luminescence \cite{fusella_plasmonic_2020}, strong-field driven nanoscale currents \cite{dombi_strong-field_2020}, enhanced nonlinear optical effects \cite{NaturePhoton6}, and strong-coupling quantum optics \cite{mueller_deep_2020,bogdanov_overcoming_2019}. However, a direct characterization method for the electric fields emerging from the resonantly excited nanostructure is still lacking. \\
Subsequent to its coherent driving by the light field, Landau damping, electron–electron, electron–phonon, and electron–surface scattering result in an ultrafast (about 10~fs) plasmon decay with the energy transferred to highly-excited, nonequilibrium carriers \cite{wang_high-q_2021,li_correlation_2020,santiago_efficiency_2020,besteiro_fast_2019,boriskina_losses_2017,dominik}. These hot carriers can contribute to chemical transformations on the NP surface, intensely studied in the field of heterogeneous catalysis \cite{zhang_surface-plasmon-driven_2018}.
The details of the plasmon decay channels, however, have only been deduced from theory and dependencies on material parameters and e.g. excitation conditions have not yet been experimentally confirmed in a direct manner \cite{avalos-ovando_temporal_2020, anderson, zentgraf}. For this reason, experiments enabling a more direct access to the plasmon decay of different NP systems are of strong interest.\\  Field sampling of plasmonic nanoantennas has been recently achieved with electro-optical sampling, limiting the approach, however, to the terahertz domain \cite{bib31}. The vast majority of plasmonic nanostructures discussed in the literature exhibits resonances in the visible spectral region, strongly motivating the extension to the petahertz (PHz) domain.
As a first example heading in this direction, plasmonic nanoantennas have been utilized as near-field sensors to enhance the sensitivity for the reconstruction of the incident light electric field ($E$-field) \cite{bib5}. What is still lacking despite its high relevance, is the realization of PHz plasmonic field sampling.\\
Here, based on recent advances in methodology \cite{Herbst_2022} we demonstrate PHz far-field sampling of plasmonic responses of colloidal NPs utilizing the tunneling ionization with a perturbation for the time-domain observation of the electric field (TIPTOE) technique \cite{bib4}. We compare the experimental data to results from finite-difference time-domain (FDTD) calculations showing that the temporal build-up and decay of the plasmon field can be resolved. Furthermore, we demonstrate extreme scale control of the transmitted fields with the NPs.
With our approach we are not only sensitive to the  electric field in the time domain, but also directly sensitive to the phase response, which is directly imprinted in the phase of the sampled field. This in turn allows for the specific design of the plasmonic material to optimize the light-matter interaction for aforementioned applications. In our case we demonstrate the ability to shape the dispersion of  ultrafast light pulses, by changes of the geometry of our NPs. Our results provide a simple way of sampling plasmon fields of arbitrary nanostructures on the PHz scale and show important practical applications in the control of light fields.\\
\begin{figure}[ht]
\centering
\includegraphics[width=\textwidth]{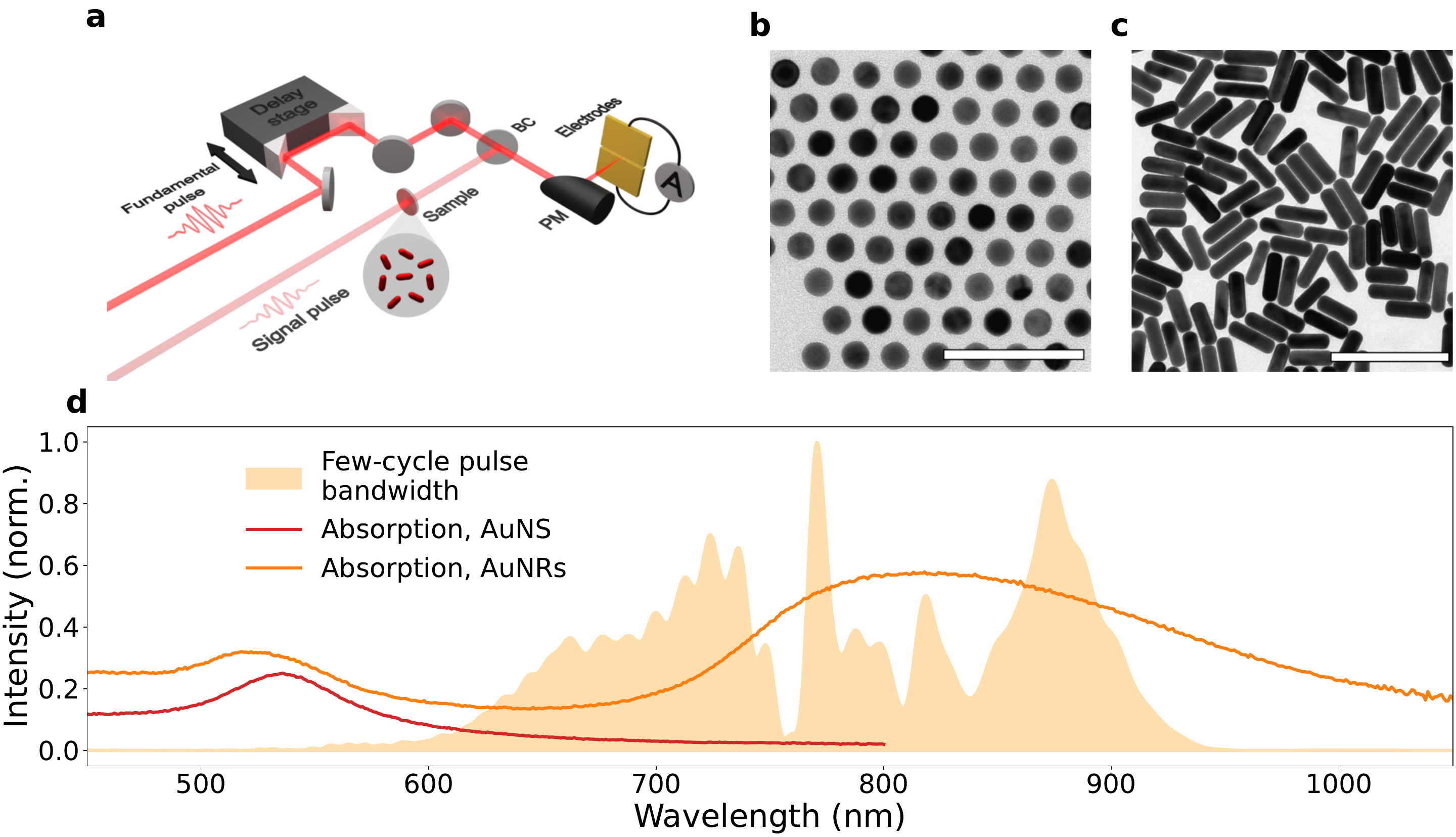}
\caption{Experimental approach. \textbf{a} Sketch of the field sampling setup.
The incident laser pulse is split into a fundamental and a signal pulse in an interferometer. The signal pulse interacts with the sample. Both pulses are recombined with variable delay and focused between two electrodes, where the TIPTOE measurement takes place. \textbf{b} TEM image of the employed AuNS with scale bar corresponding to 100~nm. \textbf{c} TEM image of the AuNRs with scale bar corresponding to 200~nm. \textbf{d} Absorption spectra of the investigated NPs (AuNS: red curve, AuNRs: orange curve) together with the incident light spectrum. BC: beam combiner, PM: parabolic mirror.}\label{fig1}
\end{figure}

\section*{Results}
\begin{figure}[ht]
\centering
\includegraphics[width=1\textwidth]{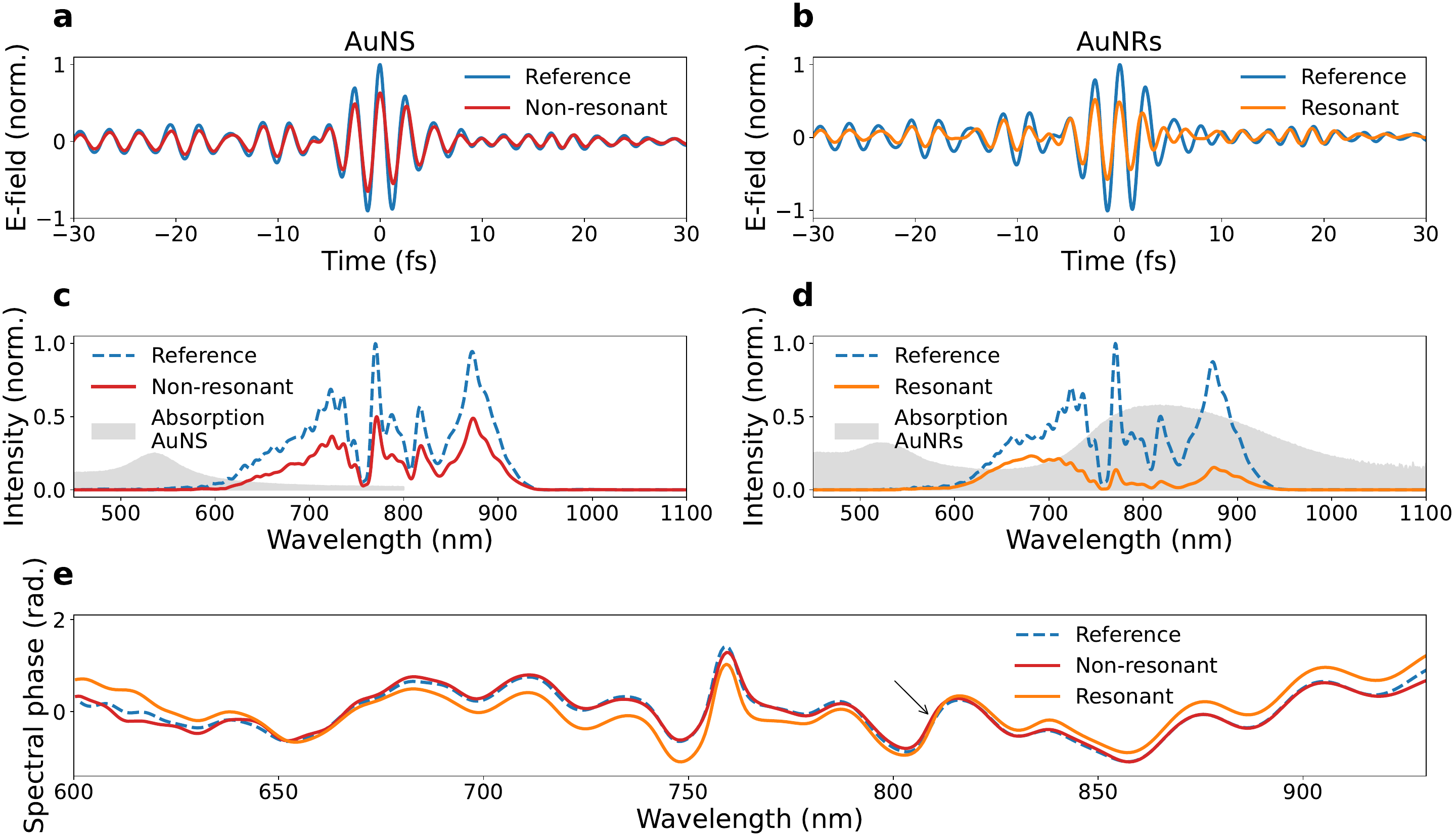}
\caption{Optical field sampling measurements. \textbf{a} TIPTOE trace for the non-resonant AuNS. \textbf{b} TIPTOE traces for the resonant AuNRs. In both cases, the blue line depicts the TIPTOE trace of the reference (plain substrate).
\textbf{c} and \textbf{d} display the corresponding transmission spectra obtained via Fourier transform from the measurements in \textbf{a} and \textbf{b}, respectively. The dashed blue curve displays the reference few-cycle spectrum. The shaded grey areas display the absorbance spectrum of the corresponding NPs. \textbf{e} Spectral phases for each case. The dashed blue line displays the reference, red the non-resonant and orange the resonant spectral phase, respectively. The black arrow indicates the crossing between the phases of the resonant and reference case.}\label{BTcomp}
\end{figure}

Few-cycle pulses with a pulse duration of 4.5\,fs, central wavelength of 780\,nm, and a repetition rate of 10\,kHz were sent into the experimental setup shown in Fig.~\ref{fig1}a. The incident light was split interferometrically into a fundamental and a signal beam. The signal beam was propagating through the plasmonic sample before being recombined with the fundamental beam for field sampling. The intensity ratio between signal and fundamental beams was chosen to be roughly 1:1000, to remain in the perturbative regime for the TIPTOE technique \cite{bib4}. The peak intensities for the excitations were chosen to be under $10^{10}~\mathrm{\frac{W}{cm^2}}$ to avoid damage of the sample, and to ensure that the interaction between the few-cycle signal pulse and the sample is in the linear regime (see Supplementary Notes 3-5). 
As samples we employed gold nanospheres (AuNS) with 20\,nm in diameter, and gold nanorods (AuNR) with dimensions of 80\,nm $\times$ 26\,nm (aspect ratio of ~3.1), respectively. Transmission electron microscopy (TEM) images of each sample are shown in Fig.~\ref{fig1}b-c. From the TEM images we also obtained the size distributions of the colloidal samples (see Supplementary Note 1). 
For AuNS, the plasmon resonance only marginally overlaps with the broad bandwidth of the few-cycle near-infrared (NIR) pulse (non-resonant case), while the longitudinal surface plasmon resonance of AuNRs lies right within the NIR spectrum (resonant case); cf. Fig.~\ref{fig1}d.
The samples were deposited onto a fused silica substrate. A replica silica substrate was used as a reference.
The dispersive contribution from the 1~mm thick fused silica substrate for the deposited NPs was carefully compensated with chirped mirrors. Observed changes should therefore only arise from the plasmon field itself.
Fig.~\ref{BTcomp} displays the results obtained from the TIPTOE measurements for both the non-resonant and resonant cases in the time domain.
As expected for a non-resonant sample, the TIPTOE measurement performed for AuNS shown in Fig.~\ref{BTcomp}a exhibits a very similar $E$-field as the reference field with a small attenuation that can be attributed to intraband transitions in gold.
This also results in almost identical spectral amplitudes and spectral phases between sample and bare substrate as shown in Fig.~\ref{BTcomp}c and e, respectively. Note that the non-resonant case serves as a benchmark in our field sampling approach.
In the case of AuNRs, a strong reduction in the amplitude of the $E$-field is observed as expected by the resonant absorption (Fig.~\ref{BTcomp}d). More interestingly, the sampled $E$-field in the time-domain (Fig.~\ref{BTcomp}b) exhibits deviations from the reference: starting from the peak of the pulse envelope we can observe a significant distortion of the optical cycles with oscillations extending in the tail of the few-cycle pulse. These distortions result in a drastic change of the spectral phase as well as shown in Fig.~\ref{BTcomp}e, which exhibits a crossing with the reference phase at the peak of the plasmon resonance around 810~nm. The phase shift due to the resonance is a well known effect and has been reported in previous studies \cite{bib5, Capasso}.

\begin{figure}[ht]%
\centering
\includegraphics[width=1\textwidth]{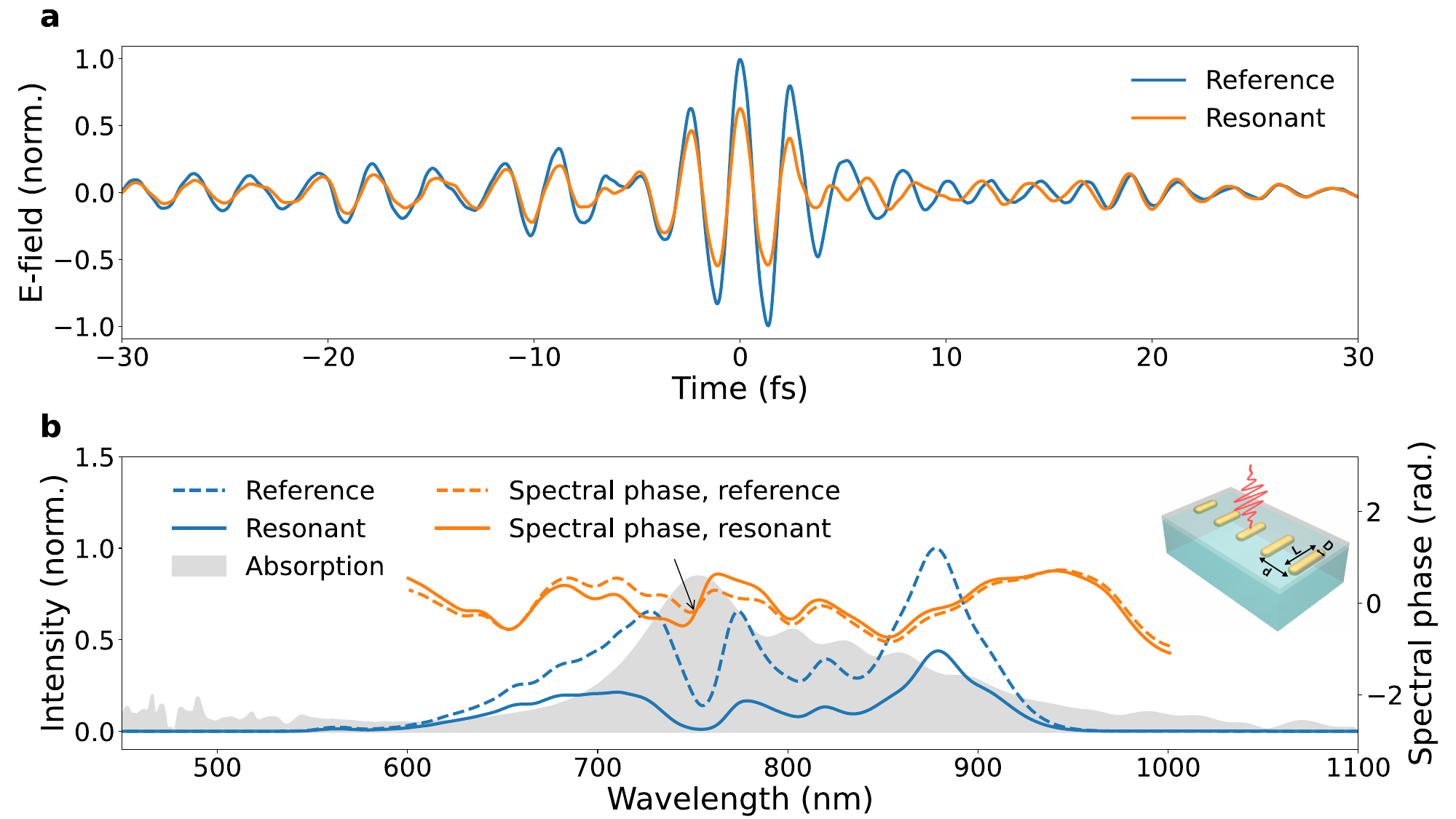}
\caption{FDTD calculation for the resonant case. \textbf{a} Simulated $E$-field with the plasmonic contribution (orange) and the input pulse as experimental data (blue). \textbf{b} Retrieved spectrum (blue) and spectral phase (orange) from the FDTD calculation, displaying the transmitted $E$-field after propagation through the plasmonic sample. As a reference the field without sample interaction is depicted in dashed lines as well. The shaded gray area displays the absorption spectrum of the sample, in which five different sized AuNRs embedded on a fused silica matrix are used for the calculations. Insets: simulation geometry of AuNRs, with five different size dimensions considering the colloidal size distribution of the AuNRs.}
\label{simulation}
\end{figure}

\noindent The experimental observations were compared to FDTD calculations implemented using Lumerical 2022 R1 software (ANSYS, Inc) \cite{alsayed2023giant}. As input parameters we used the particle dimensions, the dielectric constants of the materials and the sampled incident $E$-field. To account for the inhomogenous broadening of the plasmon resonance, we calculated the interaction between the experimental few-cycle field with five particles of different sizes, considering the determined size distribution from TEM analysis as weighting factor. The calculation results are summarized in Fig.~\ref{simulation}.

\begin{figure}[ht]%
\centering
\includegraphics[width=1\textwidth]{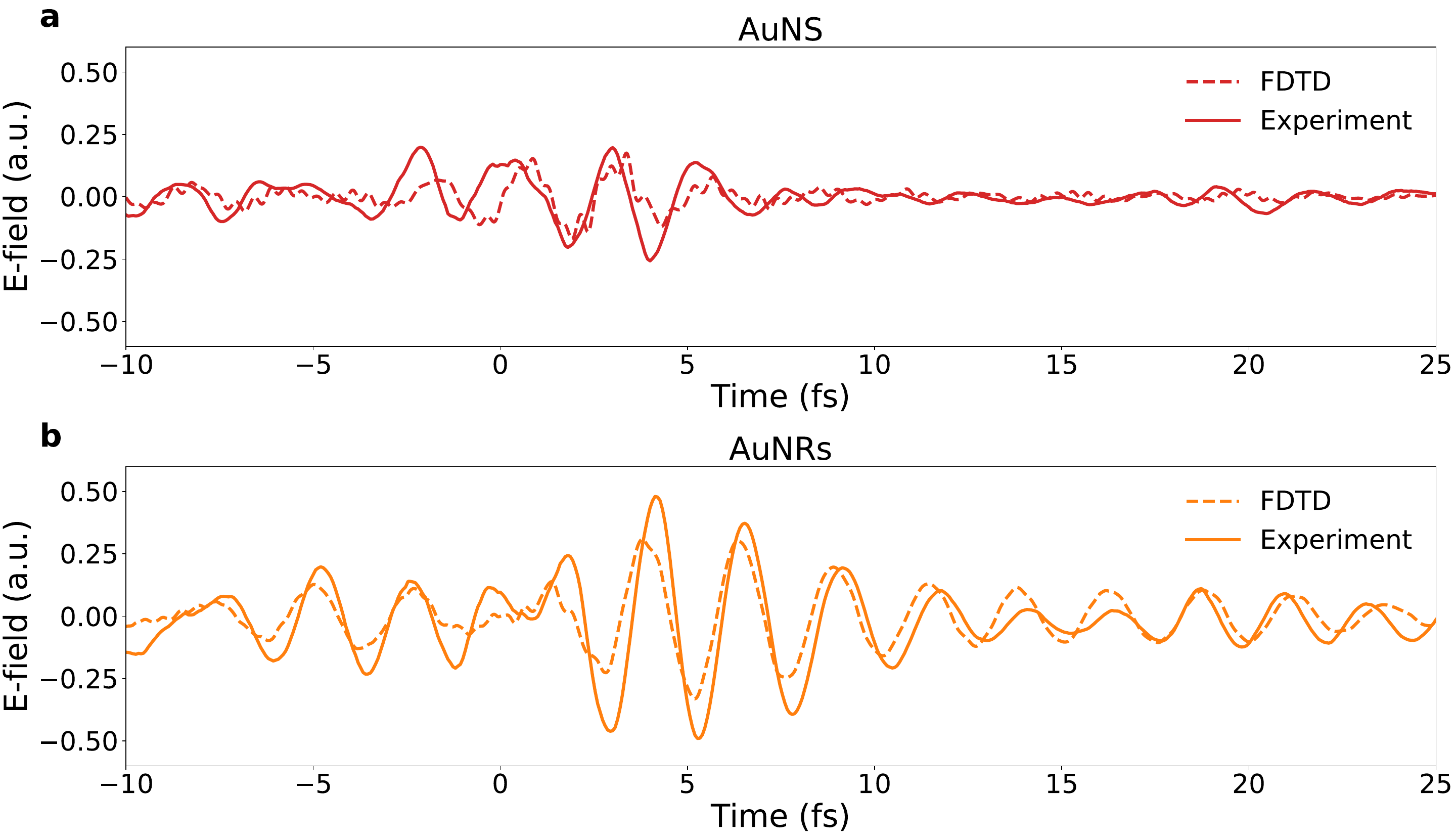}
\caption{Sampling of plasmon fields.  \textbf{a} Differential field contributions for the non-resonant AuNS with the experimental data as solid line and the FDTD data as dashed line. \textbf{b} Differential field contributions for the resonant AuNRs.}\label{BTcomp2}
\end{figure}

\begin{figure}[ht]%
\centering
\includegraphics[width=1\textwidth]{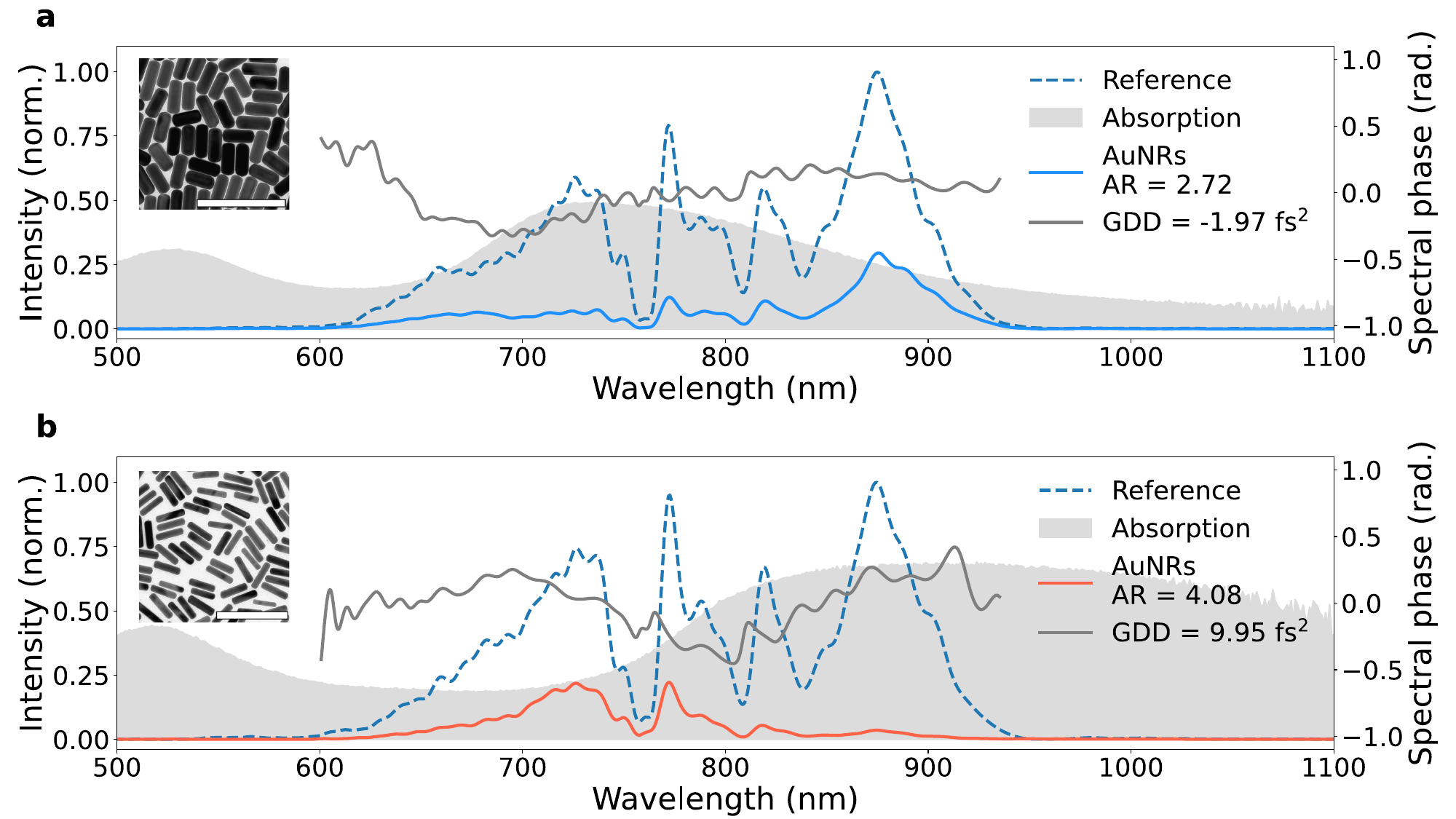}
\caption{Comparison of the propagating pulse spectral properties for AuNRs with resonances shifted relative to the AuNRs discussed before. \textbf{a} Blue-shifted AuNRs. \textbf{b} Red-shifted AuNRs. The absorption position of the plasmonic resonances are hinted as shaded grey area.The residual spectral phase contribution with its extracted GDD value is depicted as the gray line in each case. Exemplary TEM images of the AuNR samples with corresponding scale bars of 200~nm each are shown as insets.}\label{fig5}
\end{figure}

The obtained traces for the plasmonic interaction agree with the experimental traces (Fig.~\ref{simulation}a). At 5\,fs the discontinuity of the $E$-field is indicated and the delayed oscillation can be observed until 12\,fs. The agreement persists for the spectral phases. As in the experiment, the spectral phase between reference and plasmonic system exhibits a visible phase crossing at the central resonance around 750\,nm in this case, cf. Fig.~\ref{simulation}b (indicated by the arrow).\\
To resolve the onset of the plasmonic excitation from the time-resolved measurements, we subtracted the TIPTOE traces between the sample and reference. In this way, we isolate the contribution of the plasmon from the driving few-cycle field in the far-field domain. The results are reported in Fig.~\ref{BTcomp2}.
For the non-resonant case, the differential signal is almost negligible.
For the resonant case, a clear build-up of an additional field component can be resolved, with a subsequent decay with a lifetime in the order of 10\,fs, after which the differential field approaches the baseline. 
Treating the calculated data the same way, we observe a qualitative agreement with the experiment. The plasmon drive is well reproduced by the simulations, while we experimentally observe a faster decay, which could be attributed to inhomogeneous broadening due to plasmon coupling and increased damping at elevated temperatures \cite{brown_nonradiative_2016}.
The visible ultrafast decay can be attributed to the ultrafast dephasing time of the plasmon, which conversely to previous measurements \cite{bib5} can now be retrieved from far-field measurements (cf. Supplementary Note 7).
The pulse interaction with the plasmon is also reflected in the optical dispersion. Applying a polynomial fit to extract the group delay dispersion (GDD) yields different values for the plasmonic interaction compared to the reference sample. A positive GDD enhancement of approximately 8.03~fs$^{2}$ compared to the original phase value was observed after propagation through the resonant AuNRs sample. The enhancement was consistent with differently chirped pulses (see Supplementary Notes 2). Interestingly, we observe that the residual spectral phase, representing the plasmonic contribution, displays a positive parabolic shape close to the peak of the plasmon resonance. In contrast, the non-resonant AuNS induce no significant change of the dispersion. To further explore this effect, we performed measurements with samples displaying different plasmon resonances by altering the aspect ratio (AR) of the AuNRs. As samples we used AuNRs with dimensions of 71\,nm $\times$ 26\,nm and 78\,nm $\times$ 19\,nm corresponding to an AR of 2.7 and 4.1 respectively. As the AR for AuNRs changes, the longitudinal plasmon resonance shifts in frequency.
We observe that the spectral phase of the few-cycle pulse is altered correspondingly; cf. Fig.~\ref{fig5}.
As the plasmon absorption shifts relative to the previous resonant case, we also observe a shift of the positive parabolic contribution. In particular, for the plasmon resonance overlapping with the blue part of the spectrum, the minimum of the parabola shifts towards the blue, and correspondingly the remaining spectral phase exhibits a slightly negative parabolic shape that results in a negative GDD. In contrast, for the plasmon resonance overlapping with the red part of the spectrum, the minimum of the parabola shifts towards the red and results in an increase of the positive GDD. Tailoring of the plasmonic resonance is straightforward by changing the properties of the NP such as shape, size or environment and thus allows the realization of ad-hoc pulse shaping. Moreover, the ability to tailor plasmon resonances with a broad bandwidth facilitates the possibility to manipulate the spectral properties of broadband few-optical-cycle pulses. This concept was already anticipated in a theoretical study proposing the use of plasmonic NPs as metasurfaces to engineer ultrashort pulses \cite{bib6} and it has been experimentally demonstrated with 45\,fs long pulses \cite{geromel}. In this context, our approach demonstrates the feasibility of shaping broadband few-cycle pulses using broad plasmon resonances.

\section*{Conclusion}
We demonstrated the PHz sampling of resonant plasmon fields from gold nanostructures in the far-field. The implementation of the TIPTOE technique allowed direct access to the plasmon fields at visible wavelengths, which are the most relevant in nanoplasmonics. The plasmon dephasing dynamics was resolved on a subcycle scale. This will allow to benchmark theoretical descriptions and address the different channels contributing to the ultrafast dephasing. The free space integration of the samples enables investigating details of the plasmon dephasing of arbitrary plasmonic nanostructures. For example coherence in strongly coupled hybrid systems can be addressed directly \cite{saez-blazquez_can_2023,li_correlation_2020}. Furthermore, we demonstrated that the broadband plasmon resonance of nanostructures can alter the optical properties of few-cycle pulses. Additionally, the field sampling allows for direct extraction of the spectral phase without applying any reconstruction algorithms.  With careful design of such plasmonic nanostructures their usage for shaping of ultrashort laser pulses becomes in principle feasible.

\bibliography{main}

\section*{Methods}

\textbf{Sample preparation}. Monocrystalline NPs with uniform size distribution were synthesized via established wet-chemistry approaches namely the \textit{seed-mediated growth}-approach. The detailed protocol for the synthesis of AuNS is stated in ref. \cite{Xia, schulz_optimizing}. The protocol for AuNRs is stated in ref. \cite{schulz_effective_2016}. To prepare the NPs for the optical experiment, in a first step the hexadecyltrimethylammonium chloride (CTAC)/hexadecyltrimethylammonium bromide (CTAB) stabilized particles in aqueous solution were transferred to an organic solution (toluene), in which the particles are stabilized by thiol-terminated polystyrene (PSSH) with a molar weight of 25k~$\mathrm{\frac{g}{mol}}$. The reaction was carried out in 1~mL of tetrahydrofuran (THF), where the solution was stirred for three minutes in a glass vial. By shaking the vial after the reaction a thick product stuck at the side of the vial, with the remaining supernatant on the bottom, which was removed. The remained product in the vial was treated under nitrogen atmosphere to dry the sample and was then redispersed with toluene. After three washing steps via centrifugation (10k~g, 20~minutes each washing step) with toluene in which 0.1~M PSSH was dispersed, the 25~µL of the particle solution were spin-coated onto the plasma cleaned silica substrate (EKSMA Optics) at low spin speeds (100~rpms) until the organic solution evaporated completely. Materials: CTAB and CTAC were purchased from Sigma-Aldrich (USA), thiol-terminated PSSH was purchased from Polymer Source (Canada), THF ($>$99.5\%) was purchased from VWR Chemicals (USA) and toluene ($>$99.8\%) was purchased from Thermo Fisher Scientific (USA).\\
\textbf{Characterization of samples}. For the as-synthesized particles, the characterization was carried out via UV/Vis-absorption spectroscopy and transmission electron microscopy (TEM). Absorption spectra were recorded using a Varian Cary 50 spectrometer. For TEM analysis a droplet of AuNS or AuNP solution was deposited on amorphous carbon-coated copper grids. The grids were dried in air over night to remove residual solvent. TEM images were obtained using a Joel JEM-1011 transmission electron microscope operating at 100~kV The particles displayed a narrow size distribution, which was determined based on the width of the plasmon absorption and the TEM data. The deposited NPs were characterized by UV/Vis-absorption spectroscopy. For the NPs on the substrate a slight red-shift and broadening can be observed, one possible source originating from plasmon coupling. This would most likely lead to a faster damping of the plasmon oscillation. Further effects which play a role for the broadening could be changes of the environment. Nonetheless, the assumption that the observed plasmon response mainly displays the properties of the individual particles is valid, as the shift is not too pronounced.

\section*{Data and materials availability}
All data needed to evaluate the conclusions in the paper are present in the paper or the supplementary materials.

\section*{Acknowledgements}
We acknowledge fruitful discussions with Nirit Dudovich and Ferenc Krausz. This work was supported by the German Research Foundation (DFG) via the Cluster of Excellence "Advanced Imaging of Matter" (EXC 2056, 390715994). H.L. acknowledges funding by the DFG via project 432266622. M.F.K. is grateful for partial support by the Max Planck Society via the Max Planck Fellow program. J.B. acknowledges support by the Max Planck School of Photonics. M.F.K.'s work at SLAC is supported by the U.S. Department of Energy, Office of Science, Basic Energy Sciences under DE-AC02-76SF00515, and FWP SC0063. Z.W. acknowledges support from the Alexander von Humboldt Foundation. A.M.A. is grateful for support by the Researchers Supporting Project RSP-2021/152, King Saud University, Riyadh, Saudi Arabia.

\section*{Author contributions}
M.F.K. and F.C. conceived the research project and designed the experiment with H.L.; D.H. and H.L. developed the samples. K.F.W. and W.L. performed the experiments under guidance of Z.W.. A.M.A. was involved in the original design of the field sampling setup. W.L. and Z.W. performed the FDTD simulations. J.B. and T.N. supported the laser and experimental operations. V.W., E.M. and A.T. were involved in the analysis of the experimental data and assisted the preliminary characterization of the samples. K.F.W., and W.L. wrote the initial draft of the manuscript, on which all co-authors commented. 

\section*{Competing interests} None declared.

\nolinenumbers
\clearpage
\pagestyle{plain}
\setcounter{page}{1}
\setcounter{figure}{0}
\centerline{\Large\textbf{Supplementary Material on}}\vspace{.5ex}
\centerline{\Large\textbf{'Far-field Petahertz Sampling of Plasmonic Fields'}}\vspace{1ex}
\centerline{\large\textbf{by Kai-Fu Wong, Weiwei Li et al.}\vspace{2.0ex}}

\renewcommand{\thefigure}{S\arabic{figure}}

\section*{Supplementary Note 1: Determination of nanoparticle average size and aspect ratio}\label{SuppN1}

\begin{figure}[ht]%
\centering
\includegraphics[width=1\textwidth]{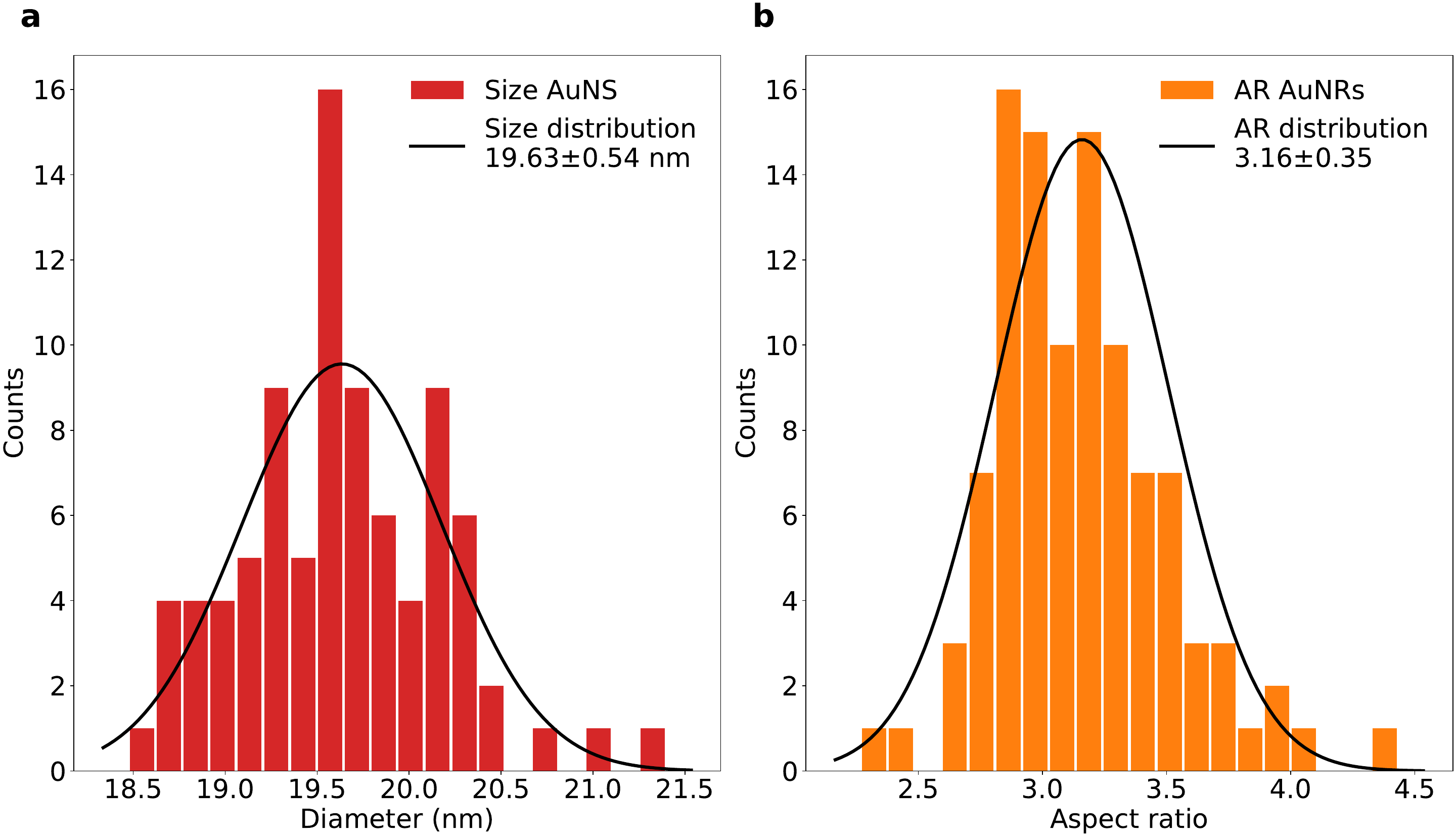}
\caption{\textbf{a} Size distribution of the employed AuNS. \textbf{b} Aspect ratio distribution of the employed  AuNRs. The black curve represents the normal distribution.}\label{dist}
\end{figure}

For the determination of the average AuNP dimensions, TEM images were analyzed. For the AuNS, the diameter is used as main quantity, while the aspect ratio (AR, length per width) serves as quantity for the characterization of the AuNRs. The determined distributions are displayed in Fig.~\ref{dist}.

From the obtained size distribution a mean diameter of 19.63~nm with a standard deviation of 0.54~nm for the AuNS is obtained. For the AuNRs a mean aspect ratio of 3.16 with a standard deviation of 0.35 is obtained. Both results are in good agreement with the values expected from the synthesis parameters of 20~nm and 3.08 respectively.

\section*{Supplementary Note 2: Extraction and validation of group delay dispersion}\label{SuppN2}

To validate the GDD values obtained from our experiments, we performed additional TIPTOE measurements by chirping the compressed few-cycle pulses using materials with defined GDD values. In our case we inserted a 1~mm thick fused silica substrate into the pathway of the signal pulse before the plasmonic interaction takes place. The results for the measurements performed in the main text while using chirped pulses are shown in Fig.~\ref{chirp_summ}:

\begin{figure}[ht]%
\centering
\includegraphics[width=1\textwidth]{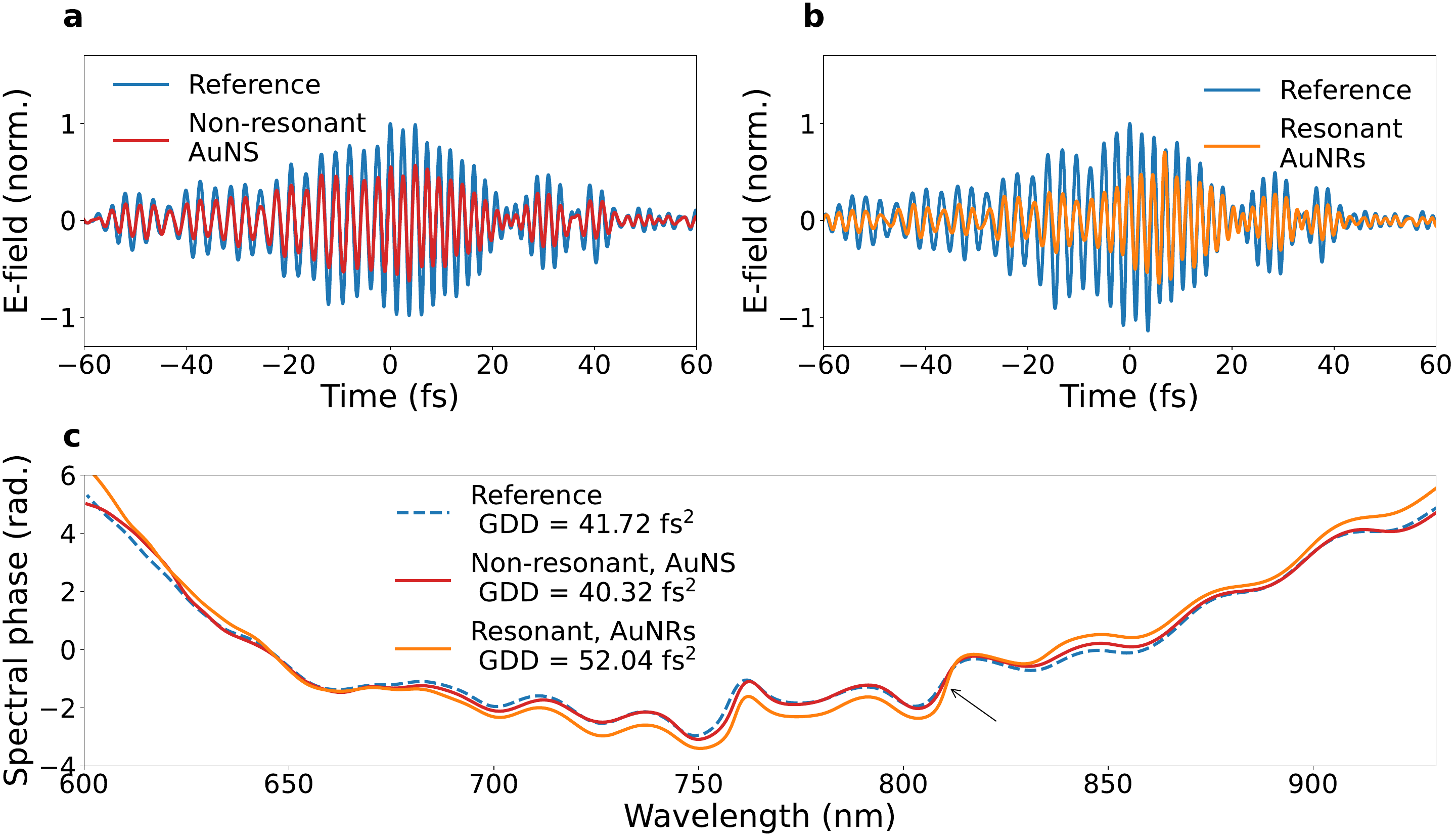}
\caption{TIPTOE traces measured for the \textbf{a} non-resonant and \textbf{b} resonant case in the time domain using chirped pulses with a known induced GDD. \textbf{c} Spectral phase of each measurement. The black arrow indicates the region of the crossing between reference and resonant spectral phase.}\label{chirp_summ}
\end{figure}

Fig.~\ref{chirp_summ} displays a clear broadening in the time- as well as in the frequency domain. Most notably the spectral phase displays a strong positive parabolic shape, indicating a positive chirp. Applying a polynominal fit to the spectral phase yields the GDD value, which for the reference is determined to be 41.72~fs$^{2}$ whereas the literature value for the GDD at 780~nm is 37.8~fs$^{2}$, which agrees well with our experimental observations. Thus, we can also assume that the deviations induced by the resonance are real, which is again confirmed by the chirped measurements, as the non-resonant case displays similar values compared to our reference as in the non-chirped case, while the resonant case shows a significant higher value compared to the reference.

\section*{Supplementary Note 3: Experimental setup and linear-response calibration}\label{SuppN3}

To obtain the few-cycle pulses required for performing the TIPTOE measurement, the output of a \SI{10}{\kilo\hertz} Ti:sapphire chirped pulse amplification (CPA) system was sent through an Argon filled hollow-core fiber, generating white light with a broad spectrum ranging from \SI{500}{\nano\meter} to \SI{950}{\nano\meter}. The white light pulses were then compressed down to \SI{4.56}{\femto\second} using chirped mirrors, and the compressed few-cycle pulses were sent into the setup for performing the TIPTOE measurements.

\begin{figure}[ht]%
\centering
\includegraphics[width=1\textwidth]{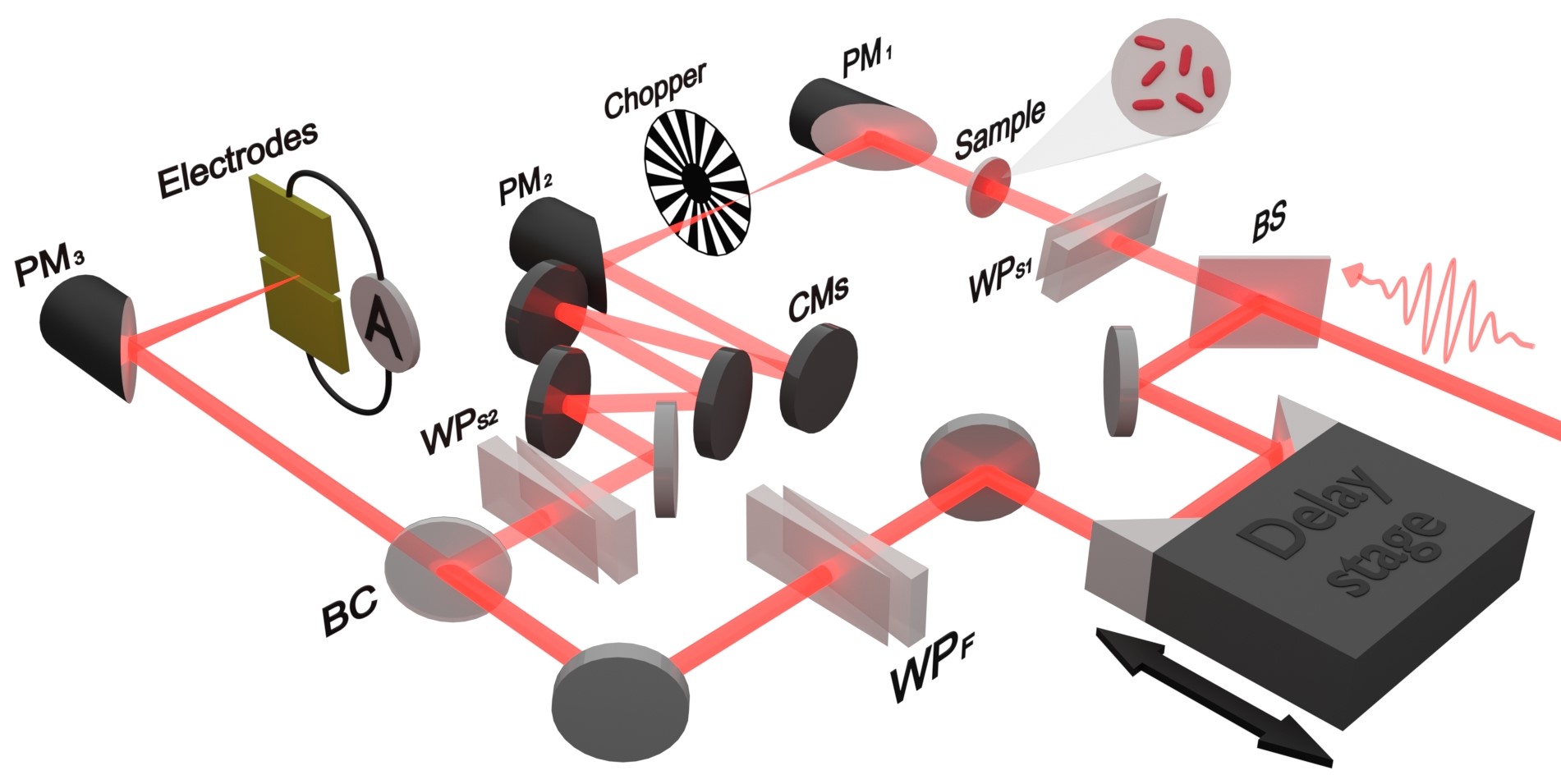}
\caption{Detailed schematic of the experimental setup.}\label{Figure S1}
\end{figure}

A sketch of the complete experimental setup is shown in Fig.~\ref{Figure S1}. The few-cycle laser pulse is split into two pathways, i.e. the strong pump beam and the weak signal beam. While the pump pulses were sent to a delay stage, the signal pulses interacted with the sample exciting the plasmon resonance. A wedge-pair (WPF) was used to finely compress the fundamental pulses which induce tunnel ionization serving as a subcycle gate for the field sampling, while another wedge-pair (WPS1) was employed to finely compress the signal pulses that interact with the gold NPs. The extra dispersive contribution from the \SI{1}{\milli\meter} thick fused silica substrate was carefully compensated with two pairs of chirped mirrors (CMs) in combination with an additional wedge-pair (WPS2). Thus, observed changes should only arise from the plasmonic field itself. Moreover, an intermediate focus was formed in the probe beam path using a pair of parabolic mirrors, where an optical chopper operating at \SI{5}{\kilo\hertz} was applied, enabling the lock-in detection of the modulated ionization yield in the form of current. Both beams were recombined and focused in between a pair of electrodes with a fixed bias of \SI{40}{\volt} in ambient environment. The transmitted probe light field which interacted with the sample can be sampled while scanning the delay stage and changing the time delay between the fundamental pulses and the signal pulses.

\begin{figure}[ht]%
\centering
\includegraphics[width=1\textwidth]{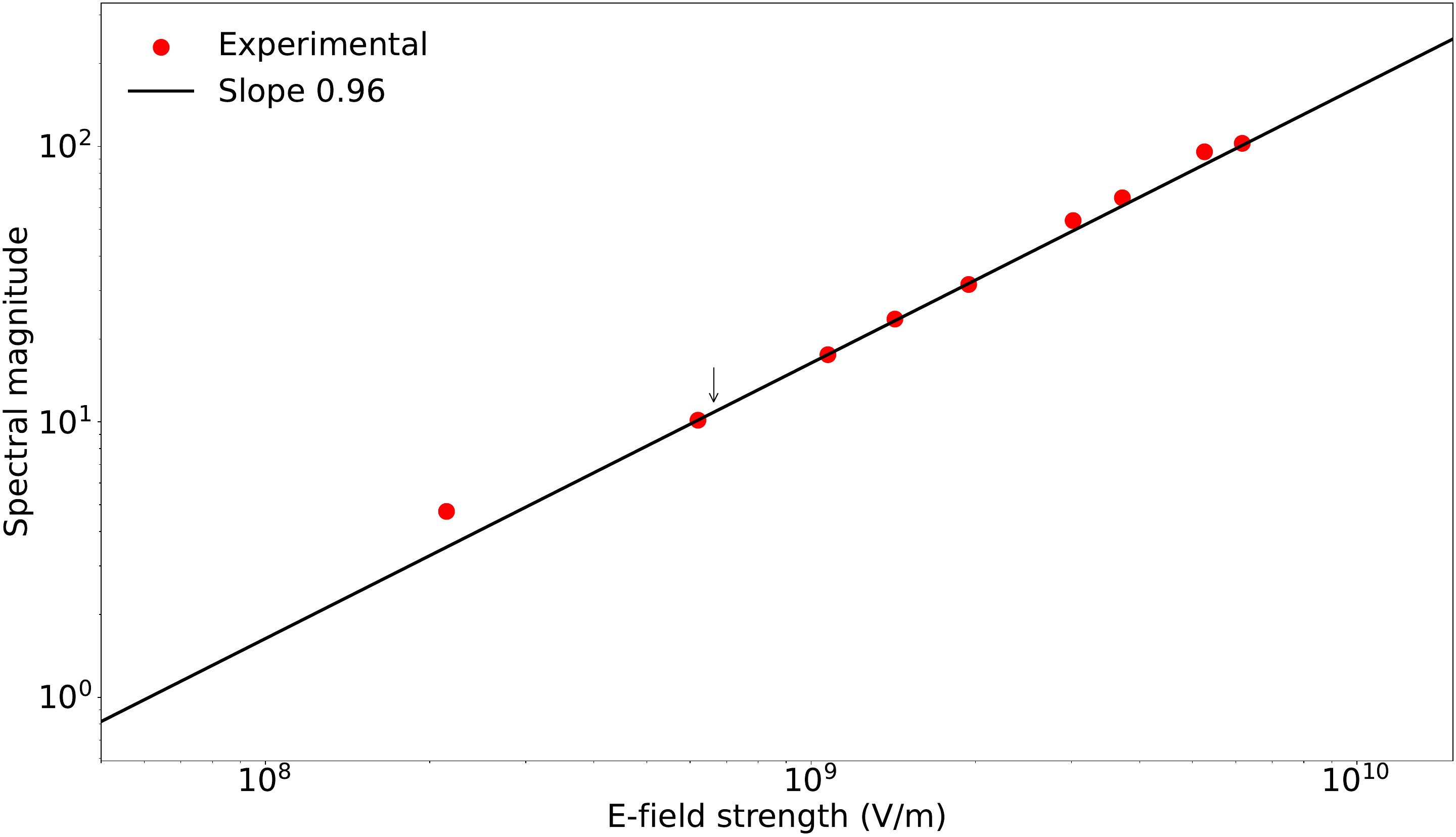}
\caption{Setup linear-response calibration: the spectral magnitudes were integrated and plotted against the $E$-field strength incident in-between the electrodes. A linear response can be identified by the slope of 0.96 in the log-log plot. The black arrow indicates the used $E$-field strength in our experiment.}\label{Figure S2}
\end{figure}

A scan of the signal $E$-field strength was done during a series of test measurements on the fused silica reference. The measured temporal traces under different $E$-field strengths incident in-between the electrodes were Fourier-transformed into the frequency domain, and the obtained spectral magnitudes were integrated in the concerned wavelength range. As can be seen in Fig.~\ref{Figure S2}, by plotting the integration against the signal $E$-field strengths, a linear response can be identified throughout the whole measured range by a linear slope in the log-log plot. This observation confirms that the setup shows a linear response to the incident signal $E$-field in a wide range, and the used signal $E$-field, as marked by the black arrow in Fig.~\ref{Figure S2}, is also weak enough to avoid inducing any unexpected influence on the measurements.

\section*{Supplementary Note 4: Linear plasmonic response determination}\label{SuppN4}

\begin{figure}[ht]%
\centering
\includegraphics[width=1\textwidth]{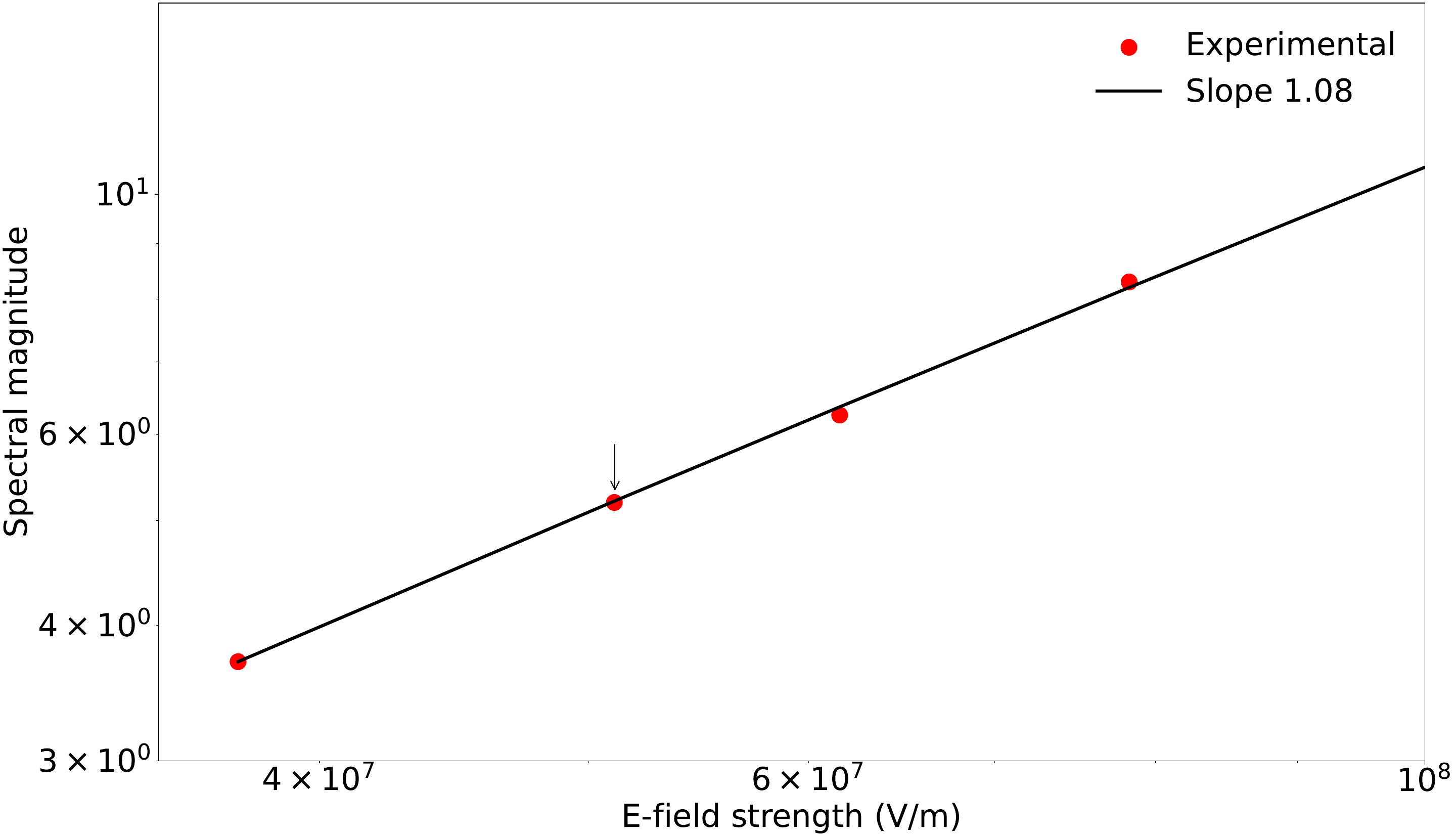}
\caption{Determination of the linear plasmonic response: the spectral magnitudes were integrated and plotted against the $E$-field strength incident onto the sample. A linear scaling of 1.08 can be seen in the investigated region in the log-log plot. The black arrow indicates the used $E$-field strength in our experiment.}\label{Figure S3}
\end{figure}

To ensure that the signal pulses are weak enough to only excite linear plasmonic response from the gold nanoparticles, a determination of the linear plasmonic response was conducted. This was done by performing the TIPTOE measurements using the resonant (AuNRs) sample with varying signal power. A Fourier transformation was applied to the measured temporal traces to obtain the corresponding spectral information in frequency domain. Similar to the process in Appendix B, in order to determine the linear response regime of the sample, the spectral magnitude profiles under different signal intensities were integrated and plotted as shown in Fig.~\ref{Figure S3}. In the linear regime the integration should in principle increase linearly with increasing signal $E$-field strength.

The plot in Fig.~\ref{Figure S3} reveals no nonlinear behavior for the system in the measured range of signal $E$-field strengths, where the typically used signal $E$-field strength was $51.1~\mathrm{{MV}\cdot{m^{-1}}}$, as marked by the black arrow. Based on this observation, the experimental measurement with AuNRs samples can most likely be assumed to happen in the regime where the plasmonic response is linear.

\section*{Supplementary Note 5: Laser pulse characterization and sample damage evaluation}\label{SuppN5}

\begin{figure}[ht]%
\centering
\includegraphics[width=1\textwidth]{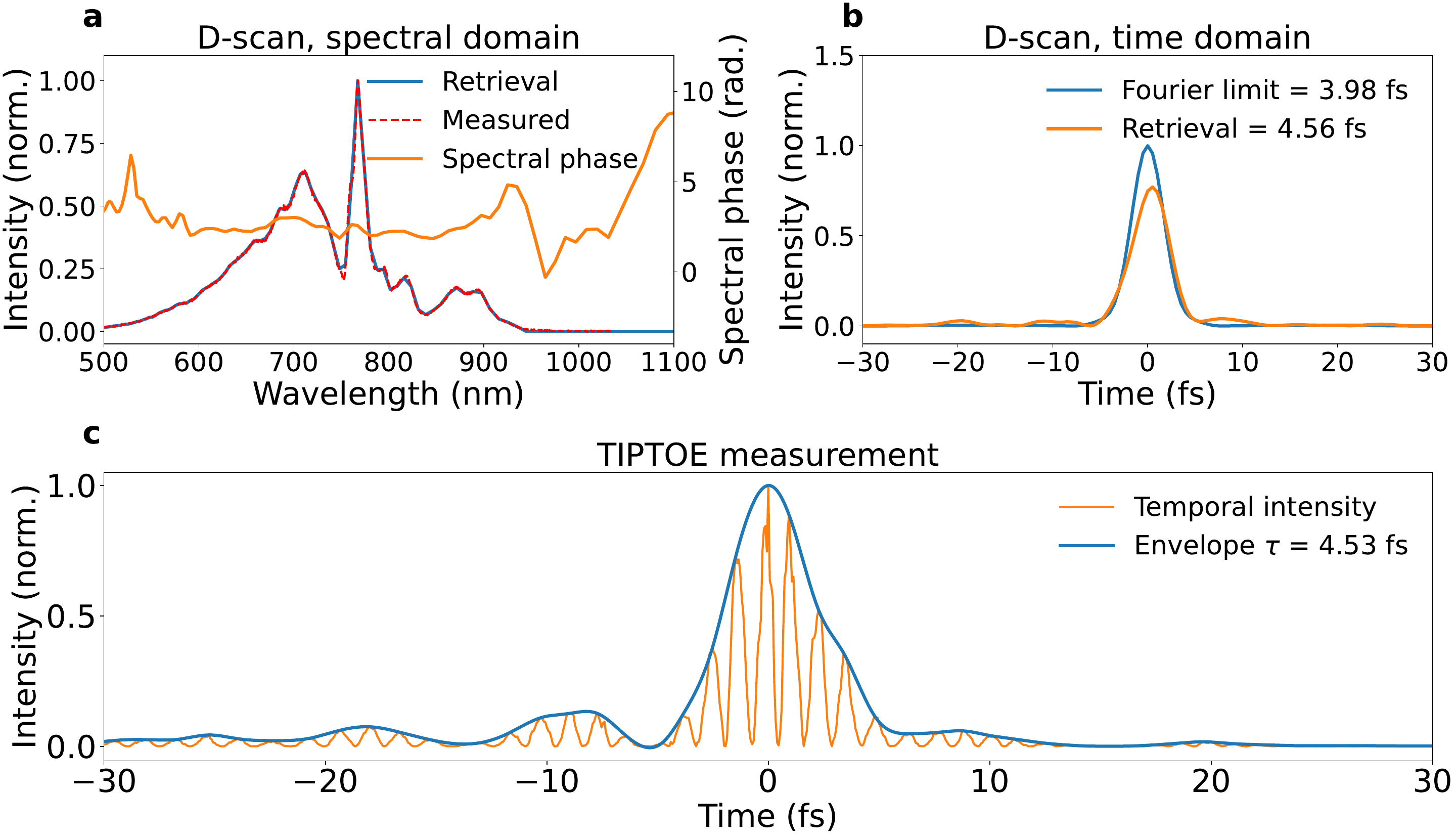}
\caption{Laser pulse characterization: \textbf{a} Reconstructed  spectrum and spectral phase. The spectrum is compared with a spectrum measured with a commerical spectrometer (dashed red). \textbf{b} Temporal results from D-scan measurement compared to the calculated transform limited pulse (solid blue). \textbf{c} Signal laser pulse obtained from the TIPTOE trace measured from the bare substrate.}\label{Figure S4}
\end{figure}

The few-cycle signal pulses used for interaction with the NPs were characterized using the dispersion-scan (D-scan) technique, and the reconstructed results are shown in Fig.~\ref{Figure S4}. The spectral intensity (blue) and phase (orange) are plotted in Fig.~\ref{Figure S4}a, where the spectrum shows a wavelength range from \SI{500}{\nano\meter} to \SI{950}{\nano\meter} with an almost flat phase curve in the region of interest, indicating a good compression of the pulses. Additionally, the reconstructed spectrum (blue) exhibits nearly the same profile as that of the measured laser spectrum (dashed red), which confirms the reliability of the D-scan measurement results. Fig.~\ref{Figure S4}b shows the reconstructed temporal intensity profile of the pulse (orange) together with the calculated Fourier Transform limited pulse (blue). The reconstructed pulse duration (full-width-half-maximum, FWHM) was measured to be \SI{4.56}{\femto\second} compared to the transform-limited duration of \SI{3.98}{\femto\second}, with a reconstruction error of 0.058\%.

Furthermore, since the TIPTOE traces in principle provide a replica of the signal laser field, information regarding the original signal laser pulse can be obtained from the TIPTOE traces measured from the bare substrate with no plasmonic contribution, as long as the induced dispersion is carefully compensated. As shown in Fig.~\ref{Figure S4}c, by taking the square of the amplitude of the measured $E$-field (orange) and taking the envelope of the positive intensity trace (blue), a laser pulse profile with a FWHM pulse width of \SI{4.53}{\femto\second} can be obtained, which is in great consistency to the results of the D-scan measurement.

\begin{figure}[ht]
\centering
\includegraphics[width=1\textwidth]{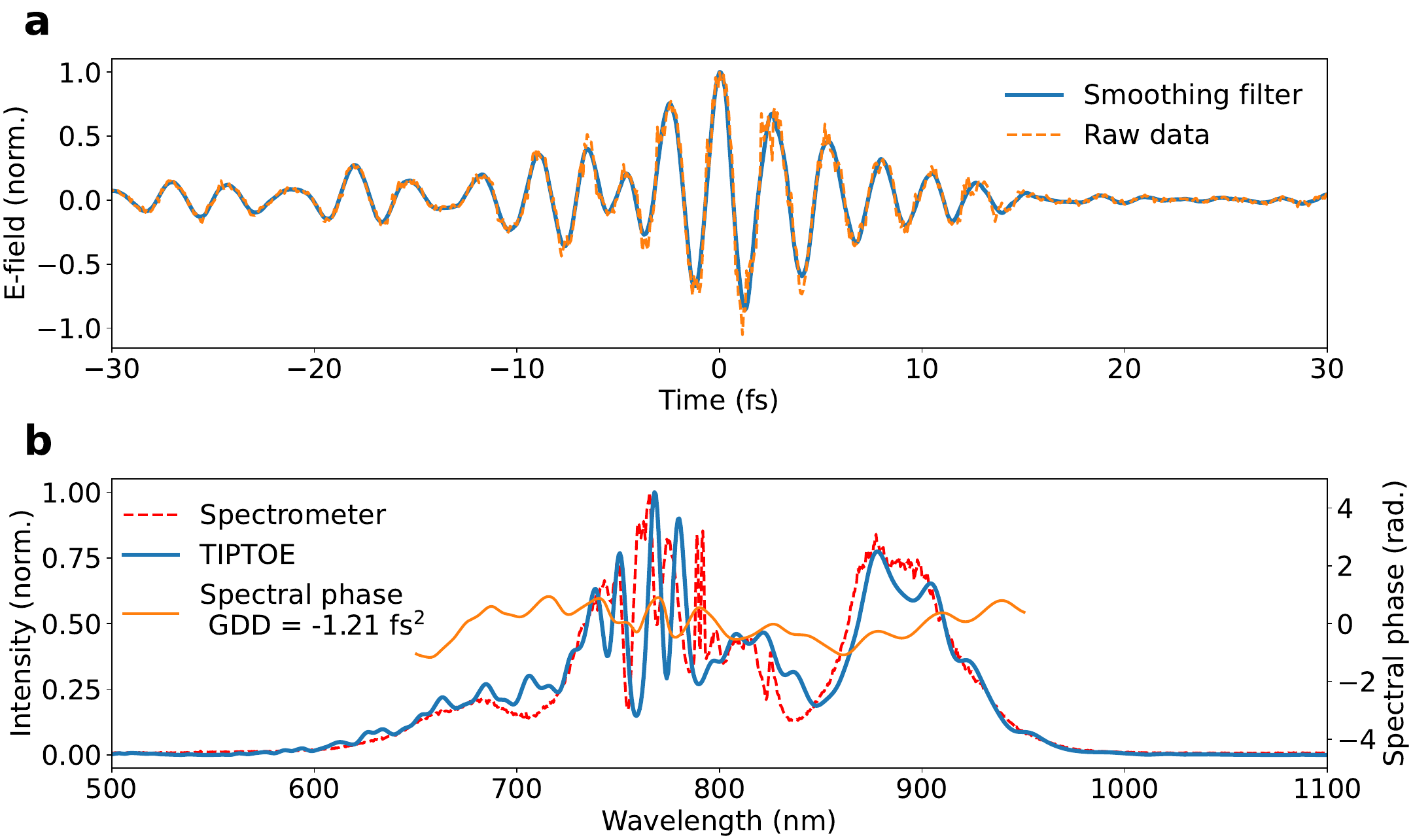}
\caption{\textbf{a} Filtered TIPTOE trace (blue) compared to the raw averaged trace (orange, dashed) obtained out of three consecutive measurements. \textbf{b} FT of the experimentally trace, which shows the spectral amplitude (blue) and the spectral phase (orange). The retrieved spectral amplitude is compared to an acquired spectrum with a commercial spectrometer (red, dashed)}\label{raw}
\end{figure}

The displayed TIPTOE traces are averaged over three consecutive measurements. To the averaged trace we applied the Savitzky-Golay-filter to filter out residual noise components. The comparison between non-filtered and filtered trace can be seen in Fig.~\ref{raw}a. To validate the accuracy of the TIPTOE measurement the FFT retrieved spectral amplitude is compared to a measured spectrum of the probe pulse under the same conditions using a commercial spectrometer, which is shown in Fig.~\ref{raw}b.

In Fig.~\ref{raw}b the retrieved spectrum from the TIPTOE compared to the spectrum obtained with the spectrometer agrees well. Additionally, the retrieved spectral phase displays a relative flat phase with a small residual negative GDD.

The damage threshold for the synthesized samples was determined by using comparable few-cycle pulses (\SI{5.9}{\femto\second}) as in the experiment at a repetition rate of 1~kHz. The range of peak intensities, which was applied to the samples ranged between $10^{10}~\mathrm{\frac{W}{cm^2}}$ and $10^{14}~\mathrm{\frac{W}{cm^2}}$. The damage was mainly determined by observation of the transmission spectrum to observe significant changes. At intensities starting from $1.5\cdot10^{12}~\mathrm{\frac{W}{cm^2}}$ we observe strong modulations of the spectrum with fringes in the few-cycle spectrum. These fringes also appear for the pure substrate, but at higher peak intensities at $2.5\cdot10^{12}~\mathrm{\frac{W}{cm^2}}$. Therefore, the determined threshold for the synthesized NPs should not exceed a value above $10^{12}~\mathrm{\frac{W}{cm^2}}$, as the range from induced damage for NPs to the fused-silica substrate itself is relatively small. We do not expect significant changes for a higher repetition rate of 10~kHz as used in the experiment, as at this frequency rate heat dissipation can still occur after two consecutive pulses. Literature which also performed damage threshold measurements on gold NPs with ultrashort laser pulses showed remarkably similar values on the same magnitude \cite{kern_damage_2011}. During our experiments we operated at intensities which were substantially lower than the determined threshold, as the highest peak intensity used in the experiment was below $10^{10}~\mathrm{\frac{W}{cm^2}}$, which means that we operated at least two magnitudes lower. This way we also ensure that we are operating in the linear regime for the interaction between the pulse and the plasmonic sample as also confirmed by the linear calibration measurements. The modulations at high intensities compared to transmission spectra below the determined damage threshold are shown in Fig. \ref{damage}:

\begin{figure}[ht]%
\centering
\includegraphics[width=1\textwidth]{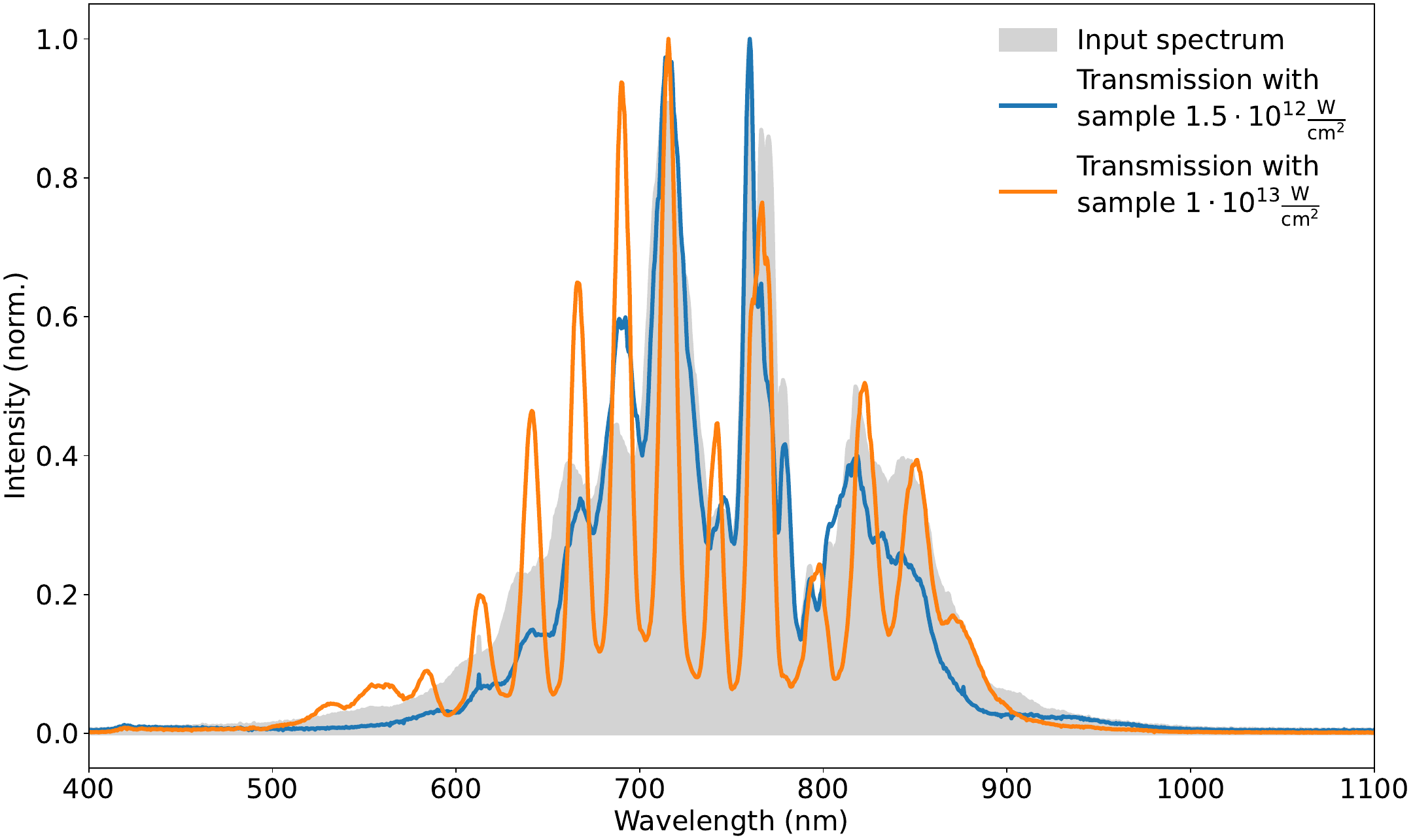}
\caption{Transmission spectra for different input intensities for the few-cycle pulse propagating through the sample. The grey shaded area represents the few-cycle transmission spectrum without any sample inserted.}\label{damage}
\end{figure}

In terms of the determined laser induced damage threshold (LIDT) value, it is ensured that the value is not exceeded for the set of experiments due to the use of relatively low power and the fact that the sample is placed in a collimated beam path. The actual laser peak intensity on the sample surface could be estimated using the following equation (\ref{damage_eq}), 

\begin{equation}
I_{0}=\sqrt{\frac{16 \ln(2)}{\pi^{3}}} \cdot \frac{W}{\tau_{\text {FWHM}} \omega_{0}^{2}}
\label{damage_eq}
\end{equation}

where $W$ is the pulse energy, $\tau_{\text {FWHM}}$ is the pulse duration, and $\omega_{0}$ is the radius of the beam. With a Gaussian beam diameter of \SI{5}{\milli\meter} on the sample and a pulse duration of \SI{4.53}{\femto\second}, the maximum peak intensity achieved with a pulse energy of \SI{0.18}{\micro\joule} is $3.464\cdot10^{10}~\mathrm{\frac{W}{cm^2}}$, which is two magnitudes below the damage threshold value determined for the AuNPs (and below the estimated value for the AuNRs as well).

\section*{Supplementary Note 6: Basic concept for data analysis}\label{SuppN6}

\begin{figure}[ht]%
\centering
\includegraphics[width=0.7\textwidth]{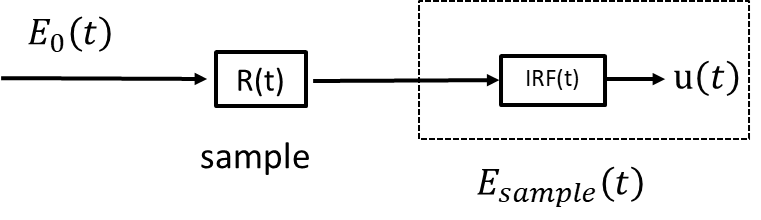}
\caption{Simplified model of the field-sampling experimental setup.}\label{LTI}
\end{figure}

Taking a bird's eye view, in the experimental scheme without sample, the lightwave detector, i.e. the TIPTOE detector, is a linear time-invariant (LTI) system when it is operated in the linear regime as shown in the dashed box in Fig.~\ref{LTI}. Therefore, the measured electric current $u(t)$ can be modeled as a temporal convolution of the instrument response function (IRF) of the TIPTOE detector and the incident light electric field $E_{0}(t)$:
\begin{equation}
{u}(t)=E_{0}(t) \ast {\textrm{IRF}}(t) \enspace .
\end{equation}

The linearity of the detector has been verified in Supplementary Note 3 (cf. Fig \ref{Figure S2}). Assuming a flat response of the detector as proven elsewhere \cite{bib4}, the measured temporal trace from the electric current is directly connected to the incident electric field as shown below:
\begin{equation}
H_{\textrm{IRF}}(\omega)=\textrm{const}  \enspace ,
\end{equation}
\begin{equation}
{\textrm{IRF}}(t) \sim \textrm{const} \ast {\delta}(t)  \enspace ,
\end{equation}
\begin{equation}
{u}(t)=E_{0}(t) \ast {\delta}(t)=E_{0}(t)  \enspace .
\end{equation}
Note that the time and frequency domain signals are connected by the Fourier transform, where the constant number, i.e. the spectrally flat response, in frequency domain is a delta function in time domain.

The linear light-matter interaction system, like in our experiment, is also a LTI system, which can be modeled as shown in Fig. \ref{Figure S3}. With the sample in beam path, the entire system can be modelled as two LTI systems put in series. It has already been demonstrated that measured electric current ${u}(t)$ in the experiment directly reflects the light electric field reaching the detector, therefore, we can safely consider the measured results as light electric field transmitted through the sample, i.e. ${u}(t) \sim E_{\textrm{sample}}(t)$.

Similarly, since it is an LTI system where its linearity has been verified in Supplementary Note 4, the measured transmitted $E$-field, $E_{\textrm{sample}}(t)$, is a temporal convolution of the impulse response function of the sample, ${R}(t)$, and the incident $E$-field $E_{0}(t)$. Considering very close values between the estimated time scale of plasmonic response and the pulse duration our incident light, a proper way of extracting the plasmonic response of the sample would be to perform a deconvolution process. In the Fourier deconvolution method, one needs to first Fourier transform $E_{\textrm{sample}}(t)$ and $E_{0}(t)$ in time domain to frequency domain $\widetilde{E}_{\textrm{sample}}(\omega)$ and $\widetilde{E}_{0}(\omega)$. The impulse response of the sample in frequency domain can then be calculated as 
\begin{equation}
\widetilde{R}(\omega)=\frac{\widetilde{E}_{\textrm{sample}}(\omega)}{\widetilde{E}_{0}(\omega)}  \enspace .
\end{equation}
Finally, the plasmon response in time domain can be obtained by inverse transforming the $\widetilde{R}(\omega)$, and $\widetilde{R}(t)={iFFT}[\widetilde{R}(\omega)]$.

Therefore, we note that while the subtraction of time traces between the transmitted and incident field as shown in the main text may not ultimately yield the plasmonic response in time domain, it can well serve to expose the distortion of the $E$-field induced by the plasmon resonance and to magnify the non-trivial contribution of the plasmonic resonances. A similar analysis was reported in \cite{bib31}.

\section*{Supplementary Note 7: Details of the FDTD simulation}\label{SuppN7}

\begin{figure}[ht]%
\centering
\includegraphics[width=12cm]{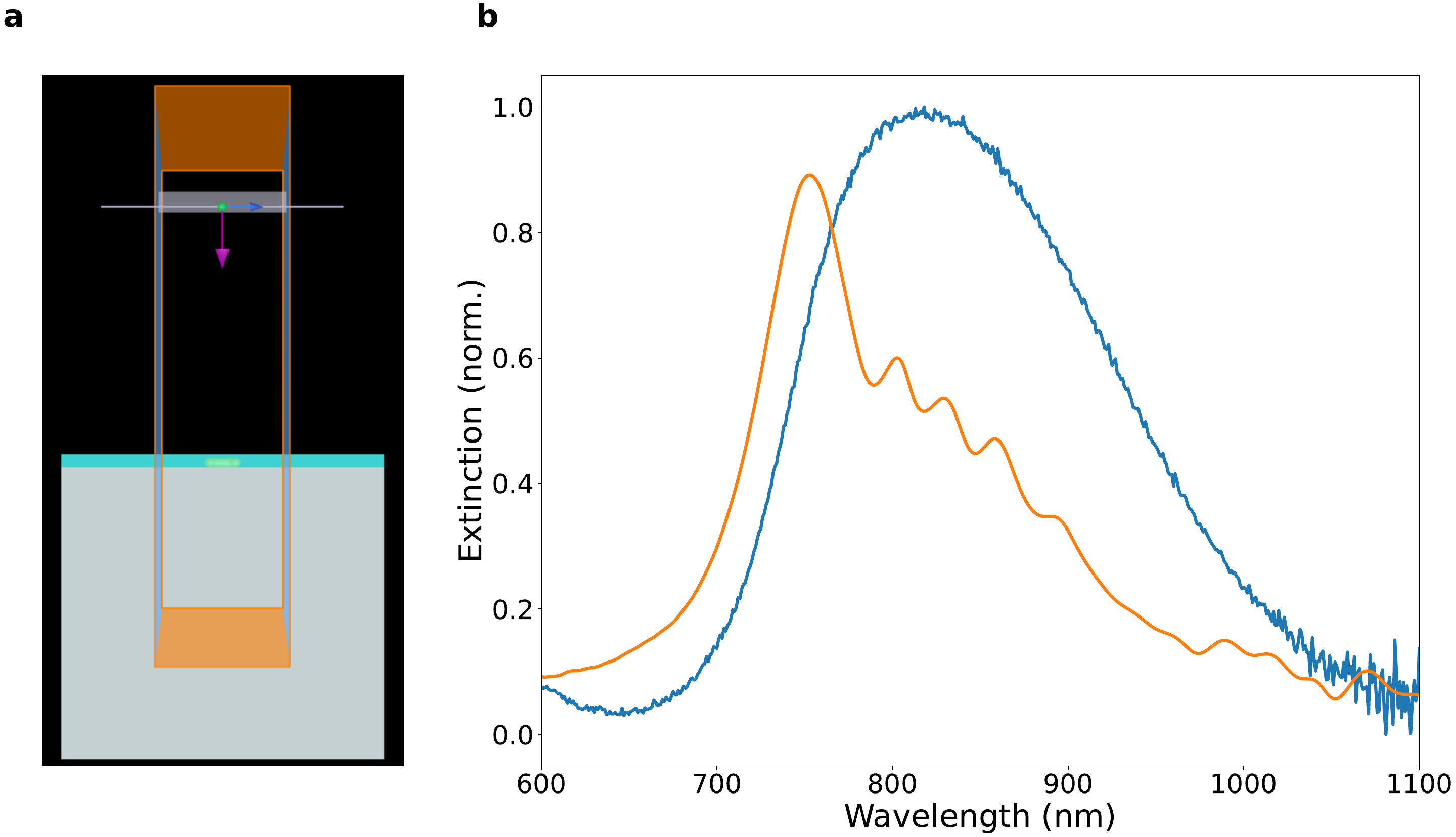}
\caption{\textbf{a} FDTD simulation model of AuNRs (green) embedded in polystyrene layer (cyan) coated on bulk fused silica substrate (gray). \textbf{b} Extinction spectrum of AuNRs sample obtained from experimental measurement (blue) and FDTD simulation results (orange).} \label{Figure S5}
\end{figure}

To theoretically determine the influence induced by the plasmonic field on the incident $E$-field, finite difference time domain (FDTD) simulations were performed using a commercial software from Lumerical. In the simulation, the dimensions of the samples with the AuNRs (AuNS as well) including the dielectric constants were given as the input parameters. The measured distinction spectra for both resonant and non-resonant cases (see Fig. 2), showing a broadband plasmonic resonance due to the fact that the prepared AuNRs and AuNS are not perfectly uniform in size. Therefore, to mimic the experimental conditions of the broadband plasmonic resonance, five different sized AuNRs with sizes considering the determined size distributions from the colloidal samples were modelled in the simulation box. Additionally, AuNS with sizes of \SI{20}{\nano\meter} in diameter were also simulated in the same condition, serving as the non-resonant reference to the resonant case. The gold NPs were embedded in a \SI{30}{\nano\meter} thick polystyrene layer coated on a bulk fused silica substrate (Fig.~\ref{Figure S5}a). To accurately model the fused silica substrate, we assigned a constant refractive index of $n=1.45$. Optical data for gold were obtained from ref. \cite{olmon2012optical}. Periodic boundary conditions were applied in the directions perpendicular to the substrate surface, while perfectly matched layer (PML) boundaries were utilized in the parallel direction. The light source used in the simulations was the incident field obtained from our experiment, which is a linearly polarized few-cycle pulse of \SI{4.6}{\femto\second} duration centered around a wavelength of \SI{780}{\nano\meter}. Time and frequency monitors are placed \SI{200}{\nano\meter} before and after the structure, ensuring a far-field detection with no influence from the gold NPs.

\begin{figure}[ht]%
\centering
\includegraphics[width=1\textwidth]{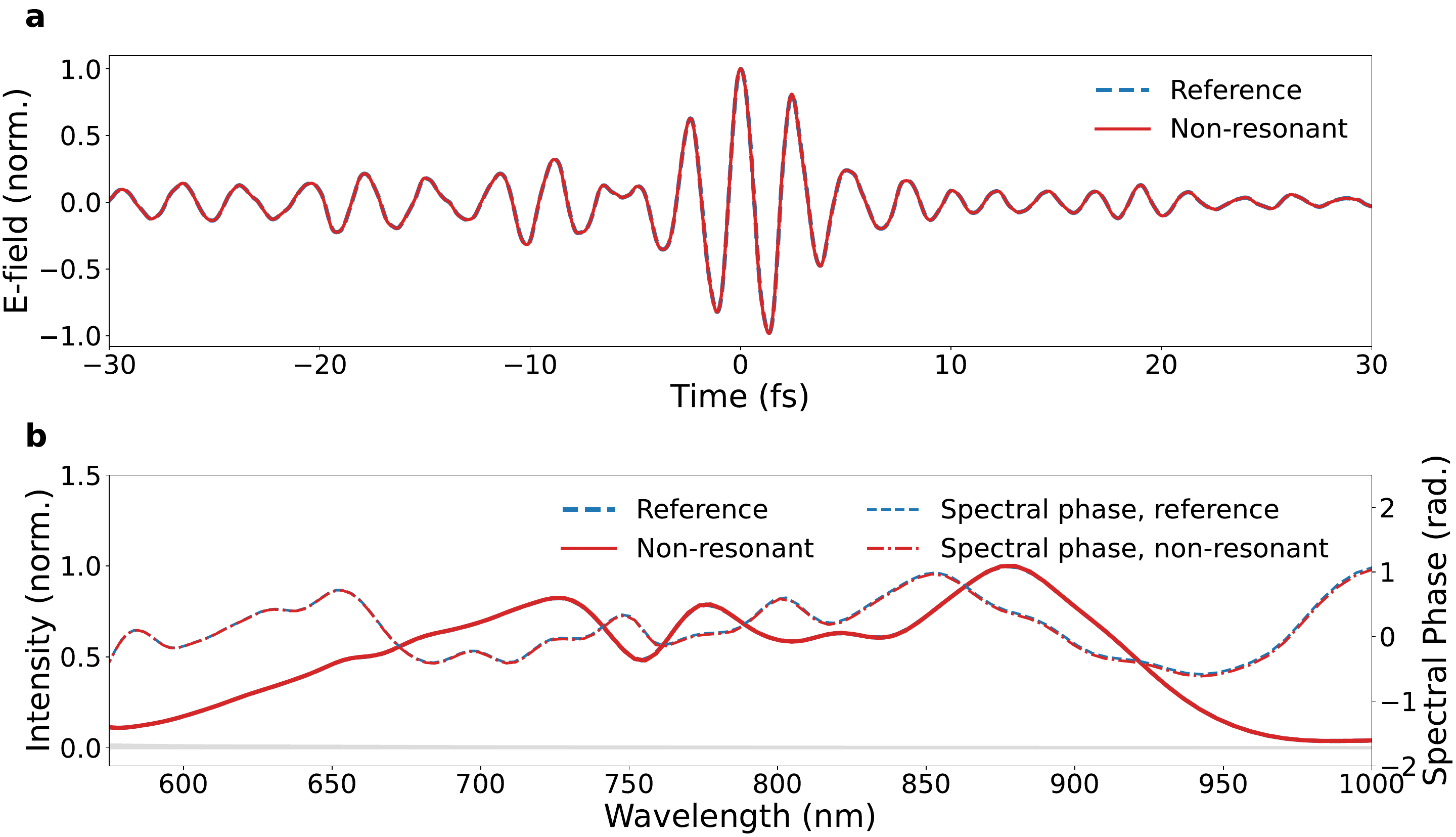}
\caption{FDTD simulation results on non-resonant AuNS. \textbf{a} Temporal traces of AuNS (red) and bare fused silica substrate (dashed blue) as reference. \textbf{b} Spectral intensity and phase of AuNS (red solid line, red dots) and bare fused silica substrate (blue solid line, blue dots) as reference, the light gray shaded area marks the absorption of the non-resonant sample.}\label{Figure S6}
\end{figure}

In the frequency domain, as shown in Fig.~\ref{Figure S5}b, for the resonant case, the bandwidth of the extinction spectrum can be well reproduced using the above-mentioned model, indicating that the simulated sample conditions are in good agreement with our experimentally investigated sample conditions. The time domain results of AuNRs (resonant case) have been shown and discussed in the main text, therefore, only the results of AuNS (non-resonant case) are shown here in Fig.~\ref{Figure S6}. Comparing to the case of AuNRs, the $E$-field interacted with AuNS does not show any obvious phase shift, and the temporal trace almost remains the same (Fig.~\ref{Figure S6}a), showing no plasmonic resonance effect from the sample. By performing a Fourier transform of the temporal traces of the incident field and the detected field, one can see that both the spectral amplitude and the spectral phase curves are almost identical (Fig.~\ref{Figure S6}b), further confirming that the observed plasmonic dephasing and phase crossing for the resonant case can not be detected in non-resonant AuNS. These results are quite reliable and consistent to the experimental observations.

\begin{figure}[ht]%
\centering
\includegraphics[width=1\textwidth]{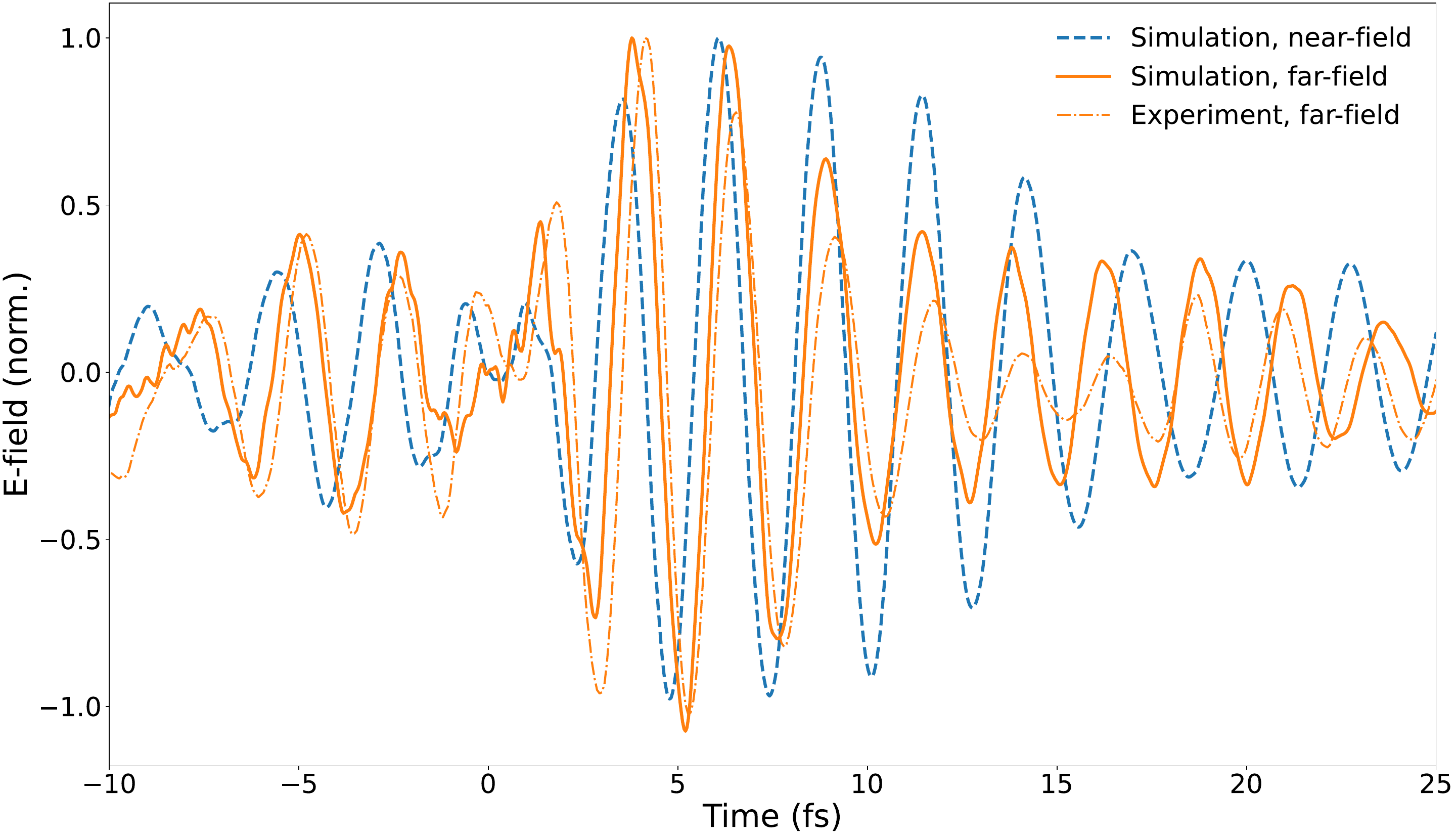}
\caption{Simulated differential plasmon fields in the near- (blue) and far-field (orange) of the resonant AuNRs. The data was treated the same way as for Fig. 2 of the main text. For comparison, the experimental data is plotted in dashed orange (cf. Fig. 2).}\label{Figure S7}
\end{figure}

In addition, a comparison between the simulated near- and far-field differential signal of the AuNRs was conducted. The results are displayed in Fig.~\ref{Figure S7}.
All data show a qualitative agreement in the first few oscillations after the driving electric field maximum. As expected the decaying window for the near-field differential is much longer than for the far-field conditions (cf. main text).

\section*{Supplementary Note 8: Time-frequency analysis}\label{SuppN9}

\begin{figure}[ht]%
\centering
\includegraphics[width=1\textwidth]{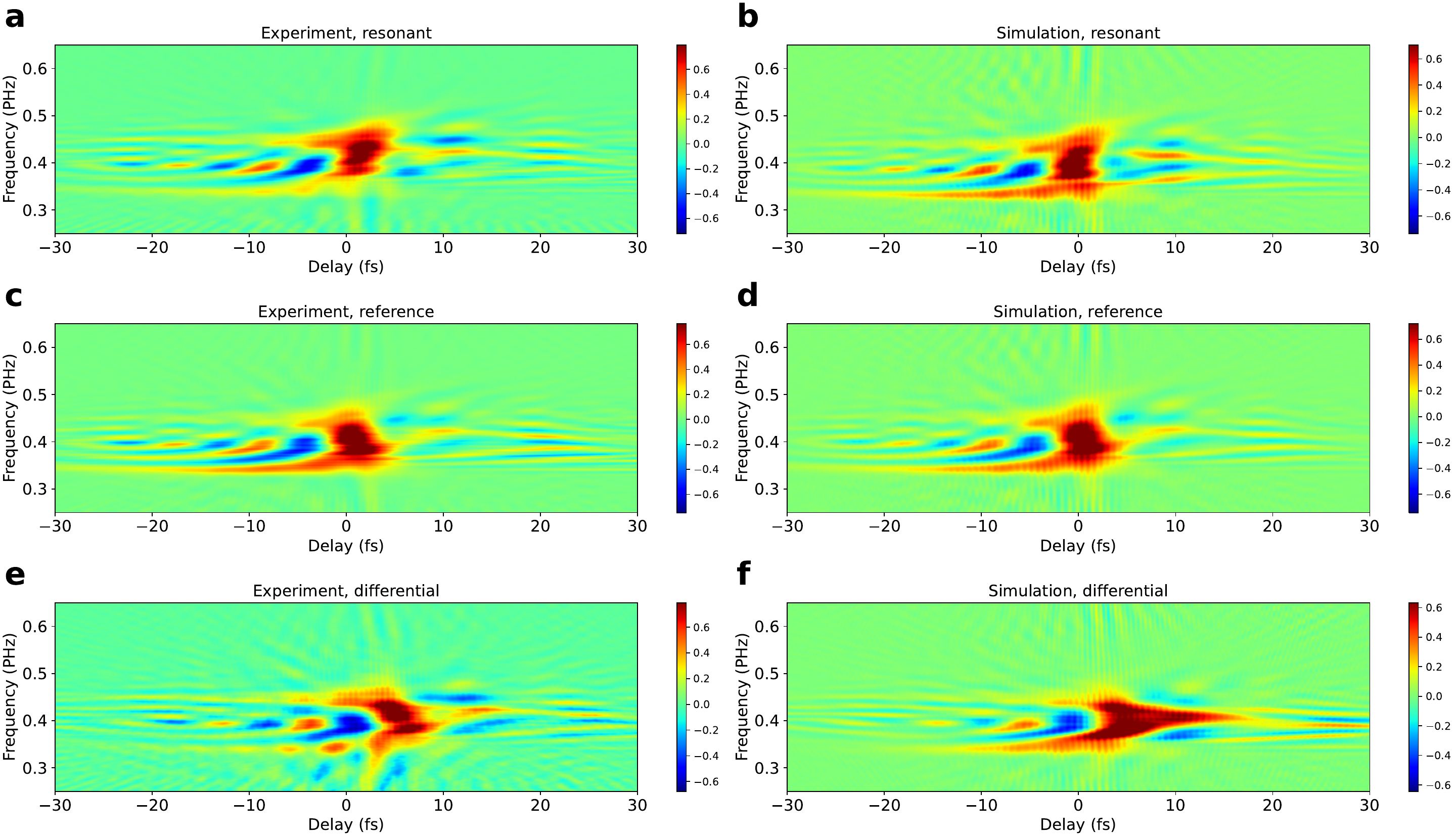}
\caption{WVD of the traces shown in the main manuscript. In the left panel the experimental traces are summarized, whereas the right panel depicts the FDTD results. ~\ref{Figure S8}\textbf{a} and \textbf{b} contain the sampled field with the resonant NPs. ~\ref{Figure S8}\textbf{c} and \textbf{d} contain the sampled field with the reference substrate. The computed difference of both traces yields the differential signal shown in ~\ref{Figure S8}\textbf{e} and \textbf{f}, which corresponds to the plasmonic response signal in the far-field domain.}\label{Figure S8}
\end{figure}

A time-frequency analysis was conducted by applying a Wigner-Ville distribution (WVD) function on the measurements, simulations and the respective differentials. The WVD response of each fields are summarized in Fig.~\ref{Figure S8} 

An agreement between experimental and simulated traces for the sample response is qualitatively observable. In the case of the reference a slight positive chirp of the red component is evident. We also observe prepulses at negative time delays, which are also indicated in the time traces as shown in the main manuscript. For the differentials we observe the buildup of the plasmonic contribution after zero delay and a chirp for both ends of the spectral frequencies as well. 

\section*{Supplementary Note 9: Influence of dielectric medium}\label{SuppN8}

\begin{figure}[ht]%
\centering
\includegraphics[width=1\textwidth]{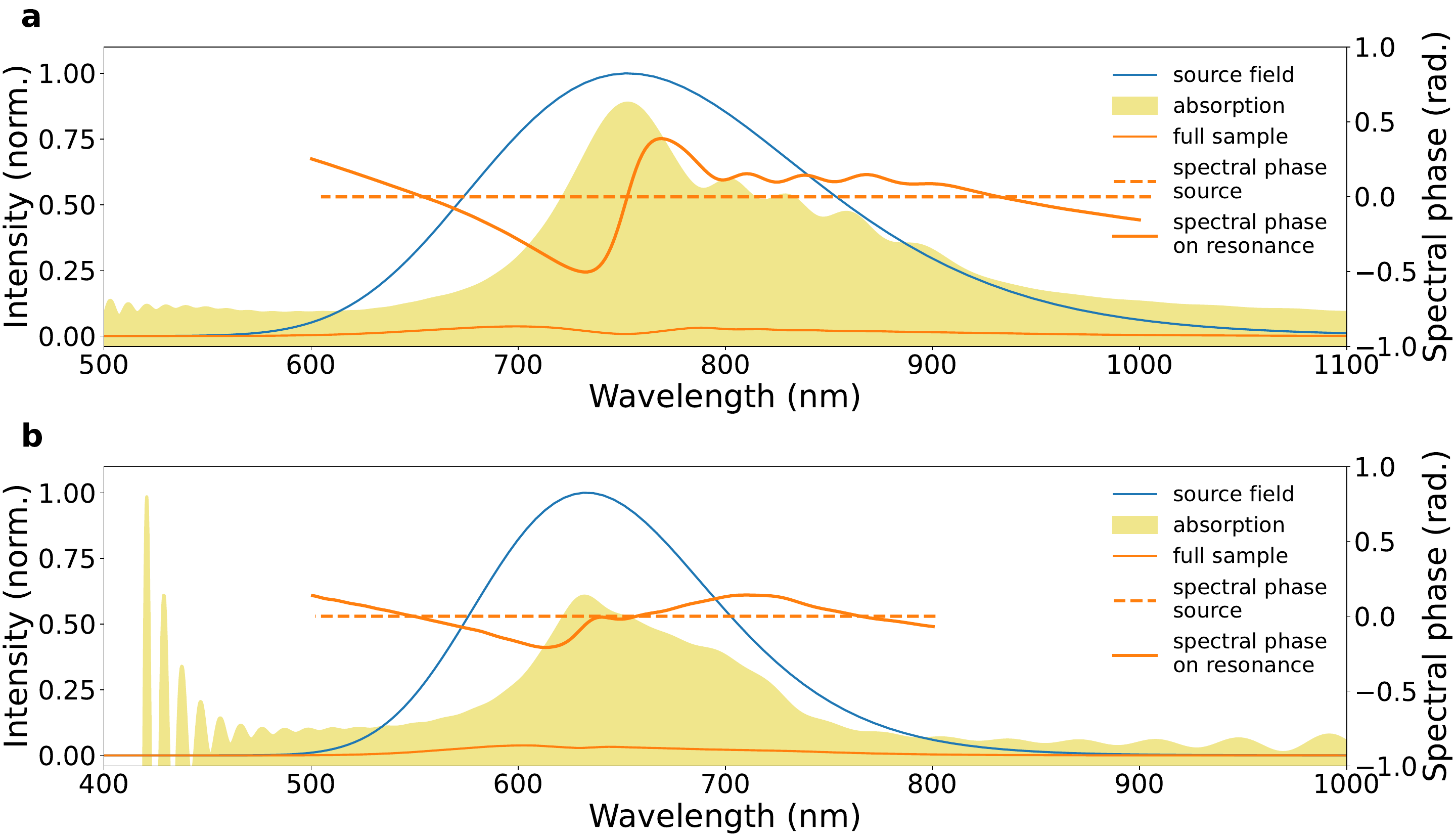}
\caption{FDTD simulations for the interaction of a simple Gaussian beam with five rods as used in the main manuscript. \textbf{a} depicts the full response with the sample as used in the initial simulation, including the polystyrene matrix. \textbf{b} displays the response without the matrix.}\label{Figure S9}
\end{figure}

To investigate the influence of the dielectric medium, which in this case is the matrix of polystyrene, in which the particles are embedded, a  FDTD simulation with and without the matrix was conducted to investigate the plasmon response depending on its environment. To account for the shift of the plasmon resonance, due to change of the medium, which leads to a change of the refractive index, we shifted our defined Gaussian pulse, with a pulse duration of \SI{4.6}{\femto\second}, accordingly on the wavelength axis and assume a negligible residual chirp due to the different dispersion response at different wavelengths. The results are shown in Fig.~\ref{Figure S9} 

The simple simulations display the influence of the dielectric medium on the plasmon resonance. We observe a strong enhancement of the plasmon resonance response in the polystyrene medium, whereas the response in vacuum, in which the particles are only attached to the substrate surface displays a lower intensity. This effect is also reflected in the spectral phase, as the phase response for the interaction with the polystyrene medium is much more pronounced. We are confident that our field sampling technique is sensitive to these changes in phase, which would prove to be advantageous in the design of metasurfaces.


\end{document}